\documentclass[prd,twocolumn,amsmath,showpacs]{revtex4}
\usepackage{graphicx}

\begin{document}

\newcommand{\Caltech}{\affiliation{Theoretical Astrophysics 350-17,
    California Institute of Technology, Pasadena, California 91125, USA}}

\title{Hybrid method for understanding black-hole mergers: Inspiralling
case} 
\author{David A.\ Nichols\footnote{Electronic address: 
\texttt{davidn@tapir.caltech.edu} } } \Caltech
\author{Yanbei Chen\footnote{Electronic address: 
\texttt{yanbei@tapir.caltech.edu} } } \Caltech
\date{\today}

\begin{abstract}
We adapt a method of matching post-Newtonian and black-hole-perturbation       
theories on a timelike surface (which proved useful for understanding head-on
black-hole-binary collisions) to treat equal-mass, inspiralling black-hole 
binaries.
We first introduce a radiation-reaction potential into this method,
and we show that it leads to a self-consistent set of equations that
describe the simultaneous evolution of the waveform and of the timelike 
matching surface.
This allows us to produce a full inspiral-merger-ringdown waveform of
the $l=2$, $m=\pm 2$ modes of the gravitational waveform of an equal-mass
black-hole-binary inspiral.
These modes match those of numerical-relativity simulations well in phase,
though less well in amplitude for the inspiral.
As a second application of this method, we study a merger of black holes with 
spins antialigned in the orbital plane (the superkick configuration).
During the ringdown of the superkick, the phases of the mass- and 
current-quadrupole radiation become locked together, because they evolve
at the same quasinormal-mode frequencies.
We argue that this locking begins during the merger, and we show that if the 
spins of the black holes evolve via geodetic precession in the perturbed 
black-hole spacetime of our model, then the spins precess at the orbital 
frequency during the merger.
In turn, this gives rise to the correct behavior of the radiation, and
produces a kick similar to that observed in numerical simulations.
\end{abstract}


\maketitle

\section{Introduction}
\label{sec:intro}

Black-hole-binary mergers are both key sources of gravitational waves
\cite{Abramovici}
and two-body systems in general relativity of considerable theoretical 
interest.  
It is common to describe the dynamics and the waveform of a quasicircular
black-hole binary as passing through three different stages: inspiral, 
merger, and ringdown (see, e.g., \cite{Flanagan}).
For comparable-mass black holes, the three stages correspond to the
times one can use different approximation schemes.
During the first stage, inspiral, the two black holes can be modeled by the
post-Newtonian (PN) approximation as two point particles
(see, e.g., \cite{Blanchet} for a review of PN theory).
As the speeds of the two holes increase while their separation shrinks, the
PN expansion becomes less accurate (particularly as the two objects begin
to merge to form a single body). 
In this stage, merger, gravity becomes strongly nonlinear (and therefore 
less accessible to approximation techniques).
After the merger, there is the ringdown, during which the spacetime
closely resembles a stationary black hole with small perturbations [and 
one can treat the problem using black-hole perturbation (BHP) theory
(see, e.g., \cite{Sasaki} for a review of BHP theory)].

Because the merger phase of comparable-mass black holes has been so 
challenging to understand analytically, there have been many attempts to 
study it with a variety of analytical tools.
One approach has been to develop PN and BHP theories to high orders in the 
different approximations.
Since neither approximation can yet describe the complete merger of black-hole 
binaries, several groups worked on developing methods that aim to get the 
most out of a given approximation technique.
The close-limit approximation (see, e.g., 
\cite{PricePullin, Gleiser, Andrade, Khanna}
for early work and \cite{Sopuerta,Sopuerta2,LeTiec,LeTiec2} for more 
recent work) and the Lazarus project (see, e.g., \cite{Baker, Campanelli2}) 
both try to push the validity of BHP to early times; the effective-one-body 
(EOB) approach (see, e.g., \cite{BuonannoDamour, BuonannoDamour2} for the 
formative work, and \cite{DamourNagar, Buonanno2, Pan, Pan2} for further developments that allow the method to replicate numerical-relativity waveforms)
aims to extend the validity of the PN approximation to later times.

There also have been several methods that do not easily fit into
the characterization of extensions of PN or BHP theories.
For example, the ``particle-membrane'' approach of Anninos et al.\
\cite{Anninos, Anninos2} computes the waveform from head-on collisions by
extrapolating results from the point-particle limit to the comparable-mass case
(and taking into account changes to the horizons computed within the membrane 
paradigm \cite{PriceThorne}).
More recently, white-hole fission was used in approximate models of black-hole 
mergers \cite{Campanelli3, Husa, Gomez}, and quite recently, Jaramillo and 
collaborators \cite{Rezzolla, Jaramillo, Jaramillo2, Jaramillo3} used 
Robinson-Trautman spacetimes as an approximate analytical model of binary
mergers (as part of a larger project correlating geometrical quantities on 
black-hole horizons with similar quantities at future null infinity).

Analytical approximations are not limited to comparable-mass 
black-hole binaries, and recently there has been a large body of work on 
developing techniques to study intermediate- and extreme-mass-ratio
inspirals (IMRIs and EMRIs, respectively).
Most of these methods aim to produce gravitational waves in ways that are 
less computationally expensive than computing the exact numerical solution
or computing the leading-order gravitational self-force are (see, e.g., 
\cite{Poisson} for a recent review of the self-force).
The majority of the approaches rely heavily on BHP techniques combined with
some prescription for taking radiative effects into account, though not
all approximate methods fall into this classification
(Barack and Cutler \cite{Barack}, for example, model EMRIs by instantaneously 
Newtonian orbits whose orbital parameters vary slowly over the orbital time 
scale because of higher-order PN effects).
A well-known example is that of Hughes \cite{Hughes}, Glampedakis 
\cite{Glampedakis}, Drasco \cite{Drasco}, Sundararajan \cite{Sundararajan} 
and their collaborators whose semi-analytical approaches are often called 
Teukolsky-based models.
These methods describe the small black hole as moving along a sequence of 
geodesics whose energy, angular momentum, and Carter constant change from the 
influence of emitted gravitational waves.
They usually involve some additional prescription to treat the transition
from the inspiral to the plunge, when the motion is no longer adiabatic.
The EOB formalism in the EMRI limit, however, does not require an assumption
of adiabatic motion (see, e.g., 
\cite{Nagar, Damour, Bernuzzi, Bernuzzi2, Bernuzzi3, Han}).
By choosing the dynamics of the EMRI to follow the EOB Hamiltonian and a 
resummed multipolar PN radiation-reaction force \cite{Damour2}, these authors 
can calculate an approximate waveform without any assumption on relative time 
scales of orbital and radiative effects.
One can also make an adiabatic approximation with EOB methods, as Yunes et 
al.\ \cite{Yunes, Yunes2} recently did in their calibration of the EOB
method to a set of Teukolsky-based waveforms.
Lousto and collaborators \cite{Lousto, Lousto2, Nakano} took a different
approach to the EMRI problem in their recent work.
They used trajectories from numerical-relativity simulations of IMRIs as a
way to calibrate PN expressions for the motion of the small black hole.
They then performed approximate calculations of the gravitational waves using 
the PN trajectories in a black-hole perturbation calculation, and found good
agreement with their numerical results.

In a previous article \cite{Nichols} (hereafter referred to as Paper I), we 
showed that for head-on collisions, one can match PN and BHP theories on a 
timelike world tube that passes through the centers of the PN theory's point
particles.
The positions of the points particles as a function of time (and, consequently,
the world tube) were chosen before evolving the waveform.
Moreover, they were selected in such a way that both PN and BHP theories were 
sufficiently accurate descriptions of the spacetime on the world tube or the 
errors in the theories did not enter into the waveform.
(A plunging geodesic in the Schwarzschild spacetime worked in Paper I.)
This allowed us compute a complete waveform for all three phases of
black-hole-binary coalescence and gave us a way to interpret the different
portions of the waveform.
Moreover, when we compared the waveform from the hybrid method with that of 
a full numerical simulation of plunging equal-mass black holes with transverse,
antialigned spins, we found very good agreement between the two.

There is no reason, {\it a priori}, why the same procedure of Paper I
(namely, specifying the position of the point particles as a function of time
and matching the metrics on a surface passing through their positions) should 
not work for inspiralling black holes as well.
The principal difficulty arises from trying to find a way of specifying the
positions of the particles for inspiralling black holes (and thus a location 
at which to match the PN and BHP metrics) that does not introduce errors into
any of the three stages of the inspiral, merger, or ringdown portions of the
waveform.
The most important development that we introduce in this paper, therefore, is 
a way of achieving this goal by including a radiation-reaction force into the 
formalism.
In the hybrid method, we compute a radiation-reaction force by using the 
outgoing waves in the exterior BHP spacetime to modify the PN dynamics in the 
interior through a radiation-reaction potential \cite{Burke}.
We show, in this formalism, that introducing a radiation-reaction potential
is equivalent to solving a self-consistent set of coupled equations that 
describe the evolution of the point particles' reduced-mass motion and the 
outgoing gravitational radiation, where the particles generate the metric 
perturbations of the gravitational waves and the waves carry away energy and 
angular momentum from the particles (thereby changing their motion).

Our principal goal in the paper is to explore this coupled set of evolution 
equations and show, numerically, that it gives rise to convergent and
reasonable results.
We will use these results to make a refinement of our interpretation of the
waveform from Paper I, and we will also compare the waveform generated by the 
hybrid method to that from a numerical-relativity simulation of an equal-mass, 
nonspinning inspiral of black holes.
The two waveforms agree well during the inspiral phase, but less well during
merger and ringdown.
The discrepancy at late times is well understood: we continue to model
the final black hole produced from the merger as nonspinning, although,
in fact, numerical simulations have shown the final hole to be spinning
relatively rapidly (see, e.g., \cite{Scheel}).
Adapting the hybrid approach to treat the final black hole as rotating is 
beyond the scope of this work, but is something that we will investigate
in the future.

As an application of the hybrid method for inspirals, we explore the large 
kicks produced from black-hole binaries with antialigned spins in the orbital 
plane (the superkick configuration \cite{Campanelli, Gonzalez}).
As noted by Schnittman et al.\ \cite{Schnittman} and emphasized to us by
Thorne \cite{Thorne}, the spins must precess at the orbital frequency during 
the final stage of the merger.
While Br\"ugmann et al.\ \cite{Brugmann} were able to replicate this
effect using a combination of PN and numerical-relativity results, we will
need to take a different approach, by using geodetic precession in the 
exterior Schwarzschild BHP spacetime, to have the spins lock to the orbital 
motion at the merger.
When we include the geodetic effect, we are able to recover the correct 
qualitative profile of the kick, although the magnitude does not match 
precisely.

We organize the paper as follows:
We review the results of Paper I in Sec.\ \ref{sec:overview},
and we describe the procedure for calculating the radiation-reaction force 
and the resulting set of evolution equations in Sec.\ \ref{sec:radrxn}.
In Sec.\ \ref{sec:numerics}, we show the convergence of our waveform,
we compare with numerical relativity, and we discuss using the hybrid
method to interpret the waveform.
Next, we discuss the behavior of spinning black holes and describe spin
precession as a mechanism for generating large black-hole kicks in Sec.\ 
\ref{sec:superkick}.
We conclude in Sec.\ \ref{sec:conclusion}.
Throughout this paper, we set $G=c=1$, and we use the Einstein summation 
convention (unless otherwise noted).

\section{A Brief Review of Paper I}
\label{sec:overview}

\begin{table*}
\begin{tabular}{cccc}
\hline
\hline
& PN spacetime & Matching shell & Perturbed Schwarzschild spacetime\\
\hline
Coordinates & $(t,R,\theta,\varphi)$ & $(t,R_s(t),\theta,\varphi)$ or
$(t,r_s(t),\theta,\varphi)$ & $(t,r,\theta,\varphi)$, $r = R+M$\\
Binary separation & $A(t)$ & $A(t)$ or $a(t)$ & $a(t)$ \\
Matching radius & $\quad R(t) = A(t)/2 \quad$ & $R_s(t) = a(t)/2 - M$ or
$r_s(t) = A(t)/2 + M$ & $r(t) = a(t)/2$ \\
\hline
\hline
\end{tabular}
\caption{(Reproduced from Paper I.)
The notation for the coordinates, the binary separation, and the matching 
radius.
We express these three variables in the PN spacetime, the BHP spacetime,
and the matching surface between the two.}
\label{notation_table}
\end{table*}

\begin{figure}
\includegraphics[width=0.95\columnwidth]{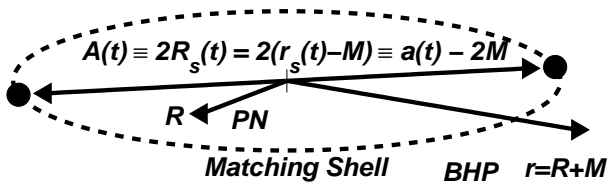}
\caption{(Reproduced from Paper I.)
The regions of spacetime and the radial coordinates in the hybrid method (at a 
given moment in time, with one spatial coordinate suppressed).
The exterior is a perturbed Schwarzschild and the interior is a PN spacetime.
At the position of the shell, the two descriptions of spacetime should both
be valid, or should be within a region of spacetime that does not heavily
influence physical observables far away.}
\label{fig:r_coords} 
\end{figure}

In this section, we will review the essentials of the formalism from Paper I.
In the hybrid method, we divide the spacetime of an equal-mass, 
black-hole-binary merger into two regions: a PN region within a spherical
shell through the centers of the PN theory's point particles, and a perturbed 
Schwarzschild spacetime outside that shell.
Figure \ref{fig:r_coords} shows this at a given moment in time (with one 
spatial dimension suppressed).
For the hybrid procedure to work, there must be either a spherical shell
on which both BHP and PN theories are simultaneously valid (to a given 
level of accuracy) or a way to prevent the errors in the 
approximations from affecting observables, such as the waveform.
By finding good agreement between the hybrid waveform and that of numerical
relativity in Paper I, we found evidence that matching the theories on a
spherical shell that passes through the PN theory's point particles works 
throughout all three stages of a head-on black-hole-binary merger: infall, 
merger, and ringdown.

To mesh the two descriptions of spacetime, we match the PN metric to that
of the perturbed Schwarzschild black hole, which involves relating the two 
coordinate systems of PN and BHP theories.
In the PN coordinate system, we will use uppercase variables, and we will
use a harmonic gauge.
For example, we will employ $(T,X,Y,Z)$ when describing the Cartesian 
coordinates of the background Minkowski space and $(T,R,\Theta,\Phi)$ 
when discussing its spherical-polar coordinates.
In the perturbed Schwarzschild spacetime, we will use $(t,r,\theta,\varphi)$,
primarily, though sometimes we will also use the light-cone coordinates,
$(u,v,\theta,\varphi)$, where 
\begin{equation}
u = t-r_* \, , \quad v=t+r_*\, ,
\end{equation}
and 
\begin{equation}
r_* = r + 2M \log\left[\frac r{2M}-1\right] \, .
\label{eq:r_star}
\end{equation}
One can match the two coordinate systems, accurate to linear order in 
$M/R$ by identifying
\begin{equation}
T=t\, ,\quad \Theta=\theta\, ,\quad \Phi=\varphi\, ,\quad R=r-M \, .
\end{equation}
For the equal-mass binaries that we study, we will denote the separation
by $A(t) = 2R(t)$ in PN coordinates and $a(t) = 2r(t)$ in Schwarzschild 
coordinates.
Moreover, because we match the two metrics on a shell passing through the 
centers of the point particles, we will indicate the position of the shell by
adding a subscript ``$s$'' to the coordinate radius.
For example, we will write $R_s(t) = A(t)/2$ or $r_s(t) = a(t)/2$ to
denote this.
For clarity, we reproduce a table that reviews the essentials of our
notation in Table \ref{notation_table}.

Because we are investigating only the lowest-order effects in our study of 
radiation reaction and large black-hole kicks, we shall only need the 
lowest-order terms in the PN metric that appeared in Paper I to describe
the interior of the shell,
\begin{eqnarray}
\nonumber
dS^2 &=& -(1-2M/R-2U_N^{(l=2)})dt^2 -8w_b^{(l=2)}dtdx^b \\
&+& (1+2M/R+2U_N^{(l=2)})(dR^2+R^2d^2\Omega) \, .
\label{eqn:PNmetric}
\end{eqnarray}
In the above equation, $M$ is the total mass of the binary, $d^2\Omega$ is
the area element on the unit sphere, $dx^b = d\theta, d\varphi$, and 
the additional variables $U_N^{(l=2)}$ and $w_b^{(l=2)}$ are the
quadrupole parts of the spherical harmonic expansion of the binary's Newtonian 
potential and gravitomagnetic potential, respectively,
\begin{eqnarray}
U_N^{(l=2)} &=& \sum_{m=-2}^2 U_N^{2,m} Y_{2,m}(\theta,\varphi)\, ,\\
w_b^{(l=2)} &=& \sum_{m=-2}^2 w_{(\rm o)}^{2,m} X^{2,m}_b(\theta,\varphi) \, .
\end{eqnarray}
We denote the scalar spherical harmonics by $Y_{2,m}(\theta,\varphi)$,
and the coefficients $U_N^{2,m}$ and $ w_{(\rm o)}^{2,m}$ are functions of $R$ 
and $t$.
The functions $X^{2,m}_b(\theta,\varphi)$ are odd-parity vector spherical
harmonics, whose $\theta$ and $\varphi$ coefficients are given by
\begin{eqnarray}
X_\theta^{l,m} &=& -(\csc \theta) \partial_\varphi Y^{l,m}(\theta,\varphi)\, ,
\\
X_\varphi^{l,m} &=& (\sin \theta) \partial_\theta Y^{l,m}(\theta,\varphi) \, .
\end{eqnarray}
A more general description is put forth in Paper I, but here we only take
the essential components needed for the calculations in the paper.

Outside of the shell, we write down a perturbed Schwarzschild metric,
\begin{eqnarray}
\nonumber
ds^2 & = & -(1-2M/r)dt^2 + (1-2M/r)^{-1}(dr^2 + r^2 d^2\Omega)\\
& & + h_{\mu\nu} dx^\mu dx^\nu,
\end{eqnarray}
where the nonzero components of the perturbed metric $h_{\mu\nu}$ that we 
shall need in this paper are the quadrupole pieces, $h_{\mu\nu}^{(l=2)}$,
and they take the form,
\begin{eqnarray}
(h_{tt}^{(l=2)})_{(\rm e)} &=& \sum_{m=-2}^2 H_{tt}^{2,m} 
Y^{2,m}(\theta,\varphi)\, ,\\
(h_{rr}^{(l=2)})_{(\rm e)} &=& \sum_{m=-2}^2 H_{rr}^{2,m} 
Y^{2,m} (\theta,\varphi)\, ,\\
(h_{\theta\theta}^{(l=2)})_{(\rm e)} &=& r^2 \sum_{m=-2}^2 K^{2,m} 
Y^{2,m}(\theta,\varphi) \, ,\\
(h_{\varphi\varphi}^{(l=2)})_{(\rm e)} &=& r^2 \sin^2\theta \sum_{m=-2}^2 
K^{2,m} Y^{2,m}(\theta,\varphi) \, ,
\end{eqnarray}
and
\begin{eqnarray}
(h_{t\theta}^{(l=2)})_{(\rm o)} &=& \sum_{m=-2}^2 h_t^{2,m} 
X_{\theta}^{2,m}(\theta,\varphi) \, ,\\
(h_{t\varphi}^{(l=2)})_{(\rm o)} &=& \sum_{m=-2}^2 h_t^{2,m} 
X_{\varphi}^{2,m}(\theta,\varphi) \, .
\end{eqnarray}
The subscripts $(\rm e)$ and $(\rm o)$ refer to the parity of the
perturbations (even and odd, respectively), where we call perturbations that
transform as $(-1)^l$ even and as $(-1)^{l+1}$ odd.

The interior PN metric must match the perturbed Schwarzschild metric on
a spherical shell between the two regions.
To make this identification, we note that because $R = r - M$, 
then the term 
\begin{equation}
\left(1-\frac{2M}r\right)^{-1} = \left(1+\frac{2M}R\right) + O[(M/R)^2] \, .
\end{equation}
We, therefore, identify the monopole piece of the PN metric with the 
unperturbed Schwarzschild metric.
Moreover, at leading order in $M/R$, we note that the perturbations of the 
two metrics match exactly,
\begin{eqnarray}
H_{tt}^{2,m} &=& H_{rr}^{2,m} = K^{2,m} = 2U_N^{2,m} \, ,\\
h_t^{2,m} &=& -4w^{2,m}_{(\rm o)} \, .
\label{perturb_matching}
\end{eqnarray}
There is then a straightforward procedure that lets one express the metric
perturbations in terms of the gauge-invariant perturbation functions of the
Schwarzschild spacetime \cite{Moncrief} (though in this paper we use the
notation of \cite{Ruiz}), which are typically called the Zerilli function and
the Regge-Wheeler function for the even- and odd-parity perturbations,
respectively.
We reproduce the expressions below:
\begin{eqnarray}
\label{eq:gauge_invariant_e}
\Psi^{2,m}_{(\rm e)} &=& \frac{2r}{3} \left\{U_N^{2,m} + \frac{r-2M}{2r+3M}
\right.\\
\nonumber
&&\times\left.\left[\left(1-\frac{2M}{r}\right) U_N^{2,m} - 
r\partial_r U_N^{2,m}\right]\right\} \, ,\\
\Psi^{2,m}_{(\rm o)} &=& 2r\left(\partial_r w^{2,m}_{(\rm o)} - \frac{2}{r}
w^{2,m}_{(\rm o)}\right) \, .
\label{eq:gauge_invariant_o}
\end{eqnarray}

In Paper I, we matched the two metrics on a timelike tube that we specified
before evolving the Regge-Wheeler and Zerilli functions.
We assumed that this tube would be spherically symmetric, and we found its
radius by first assuming the reduced-mass motion of the system followed a 
radial geodesic of a plunging test mass in the background Schwarzschild 
spacetime and then setting the radius of the world tube to be half this 
distance at each time.
This allowed us to use the PN data in the form of the Regge-Wheeler and
Zerilli functions, Eqs.\ (\ref{eq:gauge_invariant_e}) and 
(\ref{eq:gauge_invariant_o}) on this tube to provide a boundary-value
problem for the evolution of the Regge-Wheeler \cite{Regge} or 
Zerilli \cite{Zerilli} equations,
\begin{equation}
\frac{\partial^2 \Psi^{l,m}_{\rm (e,o)}}{\partial u\partial v} 
+\frac{V_{\rm (e,o)}^l\Psi^{l,m}_{\rm (e,o)}}{4} = 0 \, .
\label{wave_eq}
\end{equation}
The potentials for the Regge-Wheeler (odd-parity) or Zerilli (even-parity)
equations are given by
\begin{equation}
V_{(\rm{e,o})}^l(r) = \left(1-\frac{2M}r\right) \left(\frac{\lambda}{r^2} -
\frac{6M}{r^3}U^l_{(\rm{e,o})}(r)\right) \, ,
\label{eq:pot}
\end{equation}
where $\lambda = l(l+1)$ and
\begin{equation}
U^l_{(\rm o)}(r) = 1, \quad U^l_{(\rm e)}(r) = \frac{\Lambda(\Lambda+2)r^2+
3M(r-M)}{(\Lambda r+3M)^2} \, ,
\end{equation}
where $\Lambda = (l-1)(l+2)/2 = \lambda/2 - 1$.
After numerically solving the Regge-Wheeler or Zerilli equations above, we
computed the gravitational waveforms and the radiated energy and momentum,
all of which we found to be in good agreement with the exact 
quantities computed from numerical-relativity simulations.

In this paper, while much of the procedure we use for matching the metrics
is identical to that set forth above, there are several important differences
that we will discuss in Sec.\ \ref{sec:radrxn}.
The most important difference between the first paper and the current one 
arises in how we find the trajectory of the system's reduced mass (and then 
the timelike tube on which we match the metrics).
Before, we chose a region, prior to evolving the Regge-Wheeler and Zerilli 
equations, that would not introduce spurious effects into the results; here we 
determine the position of timelike world tube through evolving the position
of the reduced mass of the binary subject to a radiation-reaction force.
We will discuss the details of this procedure in the next section.

\section{Radiation-Reaction Potential and Evolution Equations}
\label{sec:radrxn}

In this section, we introduce a radiation-reaction potential into the hybrid
method, and we show that it leads to a set of evolution equations that
simultaneously evolve both the outgoing radiation and dynamics of the
reduced mass of the system.
This, in turn, allows us to produce a full inspiral-merger-ringdown waveform.
We first qualitatively discuss how our method works and how it compares to 
other analytical methods.
We then discuss the hybrid method in further detail, and we close this 
section by showing, analytically, that the procedure recovers 
the correct Burke-Thorne radiation-reaction potential \cite{Burke} in the 
weak-field limit.

\subsection{Qualitative Description}

\begin{figure}
\includegraphics[width=0.95\columnwidth]{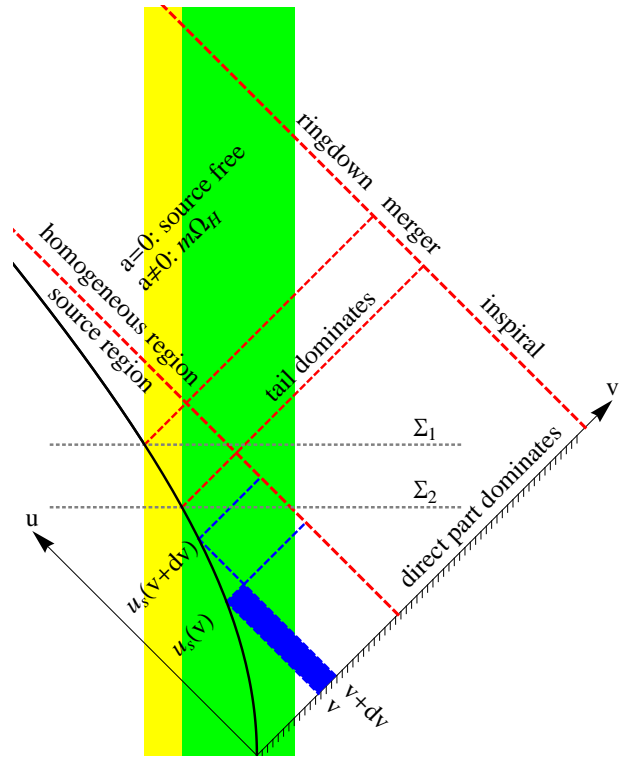}
\caption{A spacetime diagram of our method.
The solid black timelike curve depicts the region where we match PN and BHP 
spacetimes (passing through the centers of the PN theory's point particles).
Inside this curve, the spacetime can be reasonably approximated by PN theory,
whereas outside, the spacetime is better described by BHP theory.
The yellow (light gray) shade shows the strong-field region, whereas the 
green (gray) shade represents where the black-hole potential is 
weaker, but the centrifugal barrier of flat space still is important.
The dark blue (dark gray) shaded region, surrounded by the blue (dark gray) 
dashed lines shows how the value of the perturbations in the exterior (along
with the no-ingoing-wave condition) determines the value of the 
radiation-reaction potential at the next matching point.
The horizontal dashed lines represent the region of spacetime where the
close-limit approximation or Lazarus approach would begin.
The red (gray) dashed lines show how one can connect the near-zone 
behavior to the wave (through lines of constant $u$), and thereby tie the 
motion of the matching region through the black-hole effective potential to 
portions of the waveform.
This gives an interpretation of inspiral, merger, and ringdown phases
in terms of the direct and scattered parts of the waves.
Further discussion of this figure is given in the text of Sec.\
\ref{sec:radrxn}.}
\label{diagram}
\end{figure}

It is easiest to discuss our method with the aid of the spacetime diagram
in Fig.\ \ref{diagram}.
We describe the region within the solid black timelike curve with the 
near-zone PN metric, and outside this curve, we use a perturbed Schwarzschild 
region.
The black line, which passes through the PN point particles, is where we
match the two metrics.
We suppress both angular coordinates, so that each point on the curve 
represents the matching shell that we discuss in Sec.\ \ref{sec:overview}.
The shaded region represents the black-hole potential; the yellow (light gray) 
shade depicts the strong-field portion of the black-hole potential (the
strong-field near zone) and the green (gray) shade shows the region
where centrifugal potential is significant (the weak-field near zone).
There is a wave zone near the horizon (large $u$), and, consequently, the 
region where there is a large black-hole potential is confined to a small 
space in this diagram.

To have Fig.\ \ref{diagram} be an effective description of the spacetime of a
black-hole binary, both PN and BHP theories both must be sufficiently accurate 
at the PN point particles (the black line where we match the metrics), or the 
PN approximation could break down if the point particles are well-hidden 
within the black-hole effective potential.
For the errors to stay within the potential, the particles must rapidly 
fall to the horizon; thus one can see in Fig.\ \ref{diagram} that the 
black curve approaches an ingoing null ray asymptotically.
(Recall that $u$ and $v$ are the light-cone coordinates of the BHP spacetime.)
Following this trajectory, the perturbations induced by the PN spacetime
will become strongly redshifted, and they will not escape the black-hole
potential (as Price had found in his description of stellar collapse
\cite{Price}), because the potential reflects low frequency perturbations.

As in Paper I, we will again be able to interpret different portions of the
waveform by connecting a region of the waveform with the position of the
PN binary's point particles in the near zone (via constant values of the 
light-cone coordinate, $u$).
In the figure these are the thick red (gray) dashed lines of constant $u$.
The inspiral part of the waveform, which propagates directly along the light
cone, comes from the part of the trajectory within the weak-field near zone.
Once the trajectory reaches within the strong-field near zone, the waves
scatter off of the potential and propagate within the light cone (often
referred to as the PN tail part of the wave) in addition to propagating out
directly.
We view this mixed wave as characteristic of the merger phase.
Finally, as the trajectory falls within the effective potential, for a 
Schwarzschild black hole the direct part vanishes and only the scattered 
waves emerge; this part is the quasinormal ringing of the final
black hole and should be associated with the ringdown phase.
We distinguish between Schwarzschild and Kerr black holes for the ringdown
phase, because Mino and Brink \cite{Mino} and, subsequently, Zimmerman and Chen
\cite{Zimmerman} found that for Kerr black holes, frame dragging generates a
part of the waveform at the horizon frequency (that decays at a rate
proportional to the horizon's surface gravity).
This piece of the waveform looks like a source, and, thus, only when the
final black hole is not spinning do we consider the spacetime to appear
to be source free.
For this reason (and since the matching surface asymptotes to a line of
constant $v$), we call the values of $v$ greater than this limiting value the
homogeneous region, and the values of $v$ less than this the source region.

An important development in this paper is that we no longer prescribe the 
evolution of the reduced mass of the system (and thereby a matching region)
before evolving the Regge-Wheeler or Zerilli equations; rather, we specify a 
set of evolution equations for the conservative dynamics of the binary, and 
let the outgoing waves provide back reaction onto the dynamics.
This, in turn, leads to a self-consistent system of equations including
radiation reaction.
More concretely, we continue to match the PN and perturbed Schwarzschild
metrics at the centers of the PN theory's point particles.
Moreover, we will again let the reduced-mass motion of the binary system
follow that of a point particle in a Schwarzschild background; in this
paper, however, we will use the fact that there are no ingoing waves
to specify a radiation-reaction potential that acts as a dissipative force
on the Hamiltonian dynamics of the reduced mass.
This follows the spirit of the Burke-Thorne radiation-reaction potential, 
but the radiation propagates within a BHP spacetime, and, therefore, also 
takes the effects of the background curvature into account.

Furthermore, adding a radiation-reaction force to the hybrid method leads
to a set of equations that simultaneously evolve the Zerilli equation (the
waveform) and the reduced-mass motion of the binary.
In Fig.\ \ref{diagram} we represent schematically how this occurs.
We start at a given $v$ [a dark blue (dark gray) dashed line given in Fig.
\ref{diagram}] and assume that there is a no-ingoing-wave boundary condition 
along the line $u=0$.
In addition, we suppose that we have determined the black-hole-perturbation
functions for all smaller values of $v$, up to the timelike matching surface.
By evolving the Zerilli equation, Eq.\ (\ref{wave_eq}), one can find the 
Zerilli function at $v+dv$ up to the time $u_s(v)$ [within the dark blue
(dark gray) shaded region].
The no-ingoing-wave condition combined with the boundary condition on the
matching surface, however, fixes how the Zerilli function will evolve
to larger values of $u$.
When solved simultaneously with the Hamiltonian dynamics describing the
binary's motion, this lets one find the position of the reduced mass of the 
binary at $v+dv$, denoted by $u_s(v+dv)$ in the figure, and the new value of 
the Zerilli function there.
One can evolve the system for all $v$ in such a manner.

Including a radiation-reaction force does not greatly change the hybrid method 
as reviewed in Sec.\ \ref{sec:overview}.
The matching procedure works the same; one modification that comes about is 
that we must include both the Newtonian potential and the radiation-reaction
potential in the PN metric (and, therefore, gain an additional term in
the Zerilli function).
The evolution system is now quite different, because it is a coupled system of
Hamiltonian ordinary differential equations and a one-dimensional partial
differential equation.
We will discuss the system of evolution equations in greater detail after we  
compare our method with other analytical methods in the next subsection.

\subsection{Descriptive Comparison with Other Analytical and Semi-Analytical
Models}

In this section, we will compare the similarities and differences between the 
hybrid method described above and the most closely related methods mentioned 
in the Introduction: the close-limit approximation, the Lazarus program,
the comparable-mass EOB methods, the Teukolsky-based approach, the EOB
description of EMRIs, and the IMRI calculation calibrated to 
numerical-relativity data.
The comparison between the hybrid method and the other methods will be 
descriptive, but we will compare the waveform from the hybrid method with
a numerical-relativity waveform in Sec.\ \ref{sec:numerics}

To compare with the Lazarus project or the close-limit approximation, we 
again refer to Fig.\ \ref{diagram}, where we show two spacelike hypersurfaces
(the horizontal dashed lines labeled by $\Sigma_1$ and $\Sigma_2$).
In the close-limit and Lazarus methods, initial data is posed on these surfaces
at a time near the merger of the black holes.
While these approaches have been successful, posing initial data at late times
makes it more difficult to smoothly connect the initial inspiral of the binary
to the merger and ringdown later.
Moreover, because the initial data extends inside the black-hole potential,
if it contains high-frequency perturbations, these could escape the potential
barrier and enter into the waveform.
The hybrid approach escapes this problem by setting boundary data on a timelike
world tube rather than on a spacelike hypersurface.
This also lets the method connect the inspiral, merger, and ringdown portions
of the dynamics and waveform more directly.

The EOB approach, for comparable-mass ratio binaries, only describes times
prior to the merger (the hypersurfaces $\Sigma_1$ and $\Sigma_2$ in Fig.\ 
\ref{diagram}).
To create a full inspiral-merger-ringdown waveform, the EOB method
must fit a sequence of quasinormal modes to the end of the insprial-plunge
waveform.
This procedure makes a very accurate waveform, but it makes connecting the
behavior of the spacetime before and after the merger more difficult.
The hybrid method, with its interior PN region that falls toward the horizon
at late times, allows one to make a more clear connection between the dynamics
of the spacetime during inspiral and merger to that during ringdown.
In its current implementation, however, it does not produce a waveform nearly
as accurate as that of the EOB.

Although the hybrid method is designed for describing comparable-mass
black-hole binaries, it shares a few similarities and has several significant
differences from various approximate techniques that model EMRIs.
It is possible to draw a few general comparisons between the hybrid method
and the procedures for studying EMRIs, before moving to more specific 
comparisons.
While the hybrid method evolves perturbations on a black-hole background
(as most EMRI methods do), EMRI methods assume a source term as the
generator of the perturbations in the background.
The hybrid approach, however, does not have a source term; rather, the
perturbations of the background come from boundary data that correspond to
the multipolar structure of a comparable-mass PN binary.
Because the hybrid approach is a boundary-value problem, the details of
the implementation will be different from those methods that use a 
point mass as a source term.

Moving to specific EMRI models, we first compare the hybrid approach with the
Teukolsky-based methods of Sundararajan and collaborators \cite{Sundararajan}
(for example).
The hybrid approach is similar to that of \cite{Sundararajan}, in that
both use time domain codes and are capable of producing smooth
inspiral-merger-ringdown waveforms.
An important difference is that the hybrid method calculates the waveform
simultaneously with the evolution of the matching region, whereas the
EMRI method of Sundararajan computes the trajectory before the evolution
(using an adiabatic frequency-domain code during insprial, and a prescription
for the plunge and merger) and then finds the waveform from this
trajectory.
Moreover, we compute the radiation-reaction force in the hybrid method by 
matching the near-zone PN solution to an outgoing solution in the exterior
BHP spacetime, whereas the Teukolsky-based methods include radiative
effects by evolving the orbital parameters of geodesics from averaged fluxes
at infinity. 

The EOB model of Yunes et al.\ \cite{Yunes2}, is a calibration of the EOB
method to Teukolsky-based waveforms for EMRIs; it, therefore, shares the
same similarities and differences as the EOB and the Teukolsky-based methods 
discussed above.
Han and Cao \cite{Han} develop an EOB model that uses the a Teukolsky-based
energy flux (in the frequency domain) to treat radiative effects.
In comparing with the hybrid model, therefore, it also falls somewhere between
an EOB model and a Teukolsky-based method.
The recent EOB work of Bernuzzi and collaborators 
\cite{Bernuzzi, Bernuzzi2, Bernuzzi3} shares more similarity with the hybrid
method, because they evolve the Regge-Wheeler-Zerilli equations in the time
domain.
The most notable specific difference (as opposed to the general differences
between the hybrid-method and all analytical approaches to EMRIs noted above) 
is in the radiation-reaction force.
The EOB model uses a high-PN-order, resummed energy flux, whereas (as also
noted above) the hybrid method determines radiative effects from directly
matching a near-zone PN solution to an outgoing BHP solution.

We conclude this section by comparing the hybrid method with the recent
analytical work of Lousto et al.\ \cite{Lousto, Lousto2, Nakano}.
They take two approaches to calculating waveforms for IMRIs perturbatively.
In their initial work, they transform the trajectory of the small black
hole from their numerical-relativity simulations into the Schwarzschild gauge,
and they compute the waveform using this numerical trajectory in a BHP
calculation.
To be able to study a wider range of mass ratios, they use PN expressions
for the change in frequency and the radial trajectory, but use the
numerical-relativity values of the frequency to calibrate the PN
functions.
The hybrid method differs from this, because it calculates the matching
region simultaneously with the waveform, and it does not use 
numerical-relativity data to calibrate results.
Consequently, the hybrid method does not agree as well with exact results
as well as the other methods discussed here, but it does present a distinct
way of calculating the approximate spacetime and gravitational waveform.

\subsection{Radiation Reaction and Evolution Equations}

\label{sec:rr}

In this section, we will discuss the details of radiation reaction in the
hybrid method.
The end result will be the set of evolution equations described in 
Eqs.\ (\ref{eq:full1}) -- (\ref{eq:full2}), and the majority of this section
will be devoted to deriving this system of equations.

We begin, as in Paper I, with the PN metric at Newtonian order,
\begin{equation}
dS^2 = -(1-2U_N)dt^2 + (1+2U_N)(dR^2 + R^2 d^2\Omega) \, ,
\end{equation}
the same as Eq.\ (\ref{eqn:PNmetric}) of Sec.\ \ref{sec:overview}, though 
without the gravitomagnetic terms.
Here, however, we write the Newtonian potential (expanded to quadrupole
order) as
\begin{eqnarray}
\nonumber
U_N &=& U_N^{(l=0)} + U_N^{(l=2)} \\
&=& \frac M r + \sum_{m=-2}^2\left(\frac{Q_m}{R^3}Y^{2,m} 
+ F_m R^2 Y^{2,m} \right)\, .
\end{eqnarray}
The first term is the monopole piece ($M$ is the total mass of the binary) and 
the first term in the sum is the quadrupole part (and $Q_m$ are the quadrupole 
moments of the binary).
These two terms above are identical to those of Paper I, but the second
term in the sum (the polynomial in $R$ with coefficients $F_m$) is different.
One can include the terms proportional to $F_m$, because like the Newtonian
potential, they are solutions to Poisson's equation.
These terms diverge at infinity (which restricts their use to the near zone), 
but they cannot be determined from the near-zone dynamics alone, however.
Burke showed \cite{Burke}, using the technique of matched asymptotic 
expansions, that the terms with coefficients $F_m$ could represent the 
reaction of the binary in the near-zone to radiation losses to infinity.
The portion of the potential due to the moments $F_m$, therefore, is
called the Burke-Thorne radiation-reaction potential.

In the hybrid method, we will find a similar quantity in the interior PN 
spacetime by matching the PN near-zone solution to a solution in the 
Schwarzschild exterior with no ingoing waves.
Namely, when we assume that there are no ingoing waves from past-null 
infinity in the exterior BHP spacetime, this determines a radiation-reaction 
potential within the interior PN spacetime.
This allows us to incorporate the effects of wave propagation in the 
background black-hole spacetime into the dynamics of the binary.
While the Schwarzschild background does not capture every detail of the
curvature of a binary at small separation, we see that it does capture 
much of the important effects.

Proceeding with the calculation, we assume we have an equal-mass, nonspinning 
binary in the $x$-$y$ plane, located at 
\begin{equation}
{\bf X}_A(t) = -{\bf X}_B(t) = \frac 12 A(t)(\cos\alpha(t),\sin\alpha(t),0) \, ,
\label{eq:x_pos}
\end{equation}
where $A$ and $B$ are labels for the two members of the binary.
Each black hole has mass $M/2$, and a straightforward calculation shows that
\begin{eqnarray}
\label{eqQ}
Q_2(t) &=& \sqrt{\frac{3\pi}{10}}\frac{M A(t)^2}{4} e^{-2i\alpha(t)} \,,\\
Q_0(t) &=& -\frac{M A(t)^2}{4}\sqrt{\frac\pi 5} \, ,\\
Q_{-2}(t) &=& \overline Q_2(t) \,,
\end{eqnarray}
where the overline stands for complex conjugate, and where the $m=\pm 1$
components must be zero for this equal-mass binary by symmetry.
Throughout this paper, we focus just on the  $m=\pm 2$ multipoles,
because as one can see from the expressions above, the $m=0$ moment
only evolves due to the radiation-reaction force (for circular orbits), and, 
therefore, is less significant than the $m=\pm 2$ multipoles, which change on 
the orbital time scale.
Moreover, the $m=-2$ quantity is the complex conjugate of the corresponding 
$m=2$ quantity, so when we write $Q(t)$ (or any other variable that might be 
indexed by $m$), we refer to the $m=2$ variable, and similarly, for 
$\overline Q(t)$, we mean the $m=-2$ element.
This way, the notation can be simplified by dropping the $m$ label on 
multipole coefficients.
Thus, we can write the quadrupole perturbation as
\begin{equation}
U^{2,2}_N = \overline{U^{2,-2}_N}  = \frac{Q(t)}{R^3} + F(t) R^2 \, ,
\label{eq:U_N}
\end{equation}
where 
\begin{equation}
Q(t) = \sqrt{\frac{3\pi}{10}} \frac{M A(t)^2}4 e^{-i2\alpha(t)} \, ,
\label{eq:quad}
\end{equation}
and $F(t)$, an undetermined function of time, is the radiation-reaction 
potential.

One can substitute Eq.\ (\ref{eq:U_N}) into Eq.\ (\ref{eq:gauge_invariant_e}) 
and use the fact that $r=R-M$ to find the Zerilli function.
Calculating the Zerilli function introduces many factors of $M/R$ into the end
result, which, because our calculation is only accurate to Newtonian order, we 
will keep only the leading-order terms in $R$.
We find that
\begin{equation}
\label{eqZQF}
\Psi_{(\rm e)} = \frac{2Q(t)}{R^2} + \frac{F(t)R^3}{3}  \, .
\end{equation}
We will also shortly need expressions for the derivative of the Zerilli 
function with respect to the tortoise coordinate $r_*$, Eq.\ (\ref{eq:r_star}),
which we compute here as well.
Again, we will keep the leading-order expression in $R$, but we will also
retain the factor of 
\begin{equation}
\frac{dr}{dr_*}= \left(1-\frac{2M}r\right) = \frac{(R-M)}{(R+M)}\, ,
\end{equation}
since although $\Psi_{(\rm e)}$ may be constant on the horizon, 
$\partial\Psi_{(\rm e)}/\partial r_*$ should vanish there \cite{Price}.
The result of this calculation is that
\begin{equation}
\label{dpsi_drs}
\frac{\partial \Psi_{(\rm e)}}{\partial r_*} = \left(\frac{R-M}{R+M}\right)
\left(-\frac{4 Q(t)}{R^3} + F(t) R^2 \right)\, .
\end{equation}

The Zerilli function satisfies the simple wave equation in a potential,
Eq. (\ref{wave_eq}).
As before, the value of the Zerilli function at the matching surface, 
$R_s(t) = A(t)/2$, provides a boundary condition for the Zerilli equation
on the matching surface, but now there is an additional boundary condition
on the Zerilli function's derivative with respect to the tortoise coordinate.
The two boundary conditions state that
\begin{eqnarray}
\label{eqZQFb}
\Psi_{(\rm e)}(t) &=& \frac{8Q(t)}{A(t)^2} + \frac{F(t) A(t)^3}{24}\, ,\\
\label{eqdPsi_drsb}
\frac{\partial \Psi_{(\rm e)}(t)}{\partial r_*} &=& \left(\frac{A(t)-2M}
{A(t)+2M}\right)  \\
\nonumber
&& \times \left(-\frac{32Q(t)}{A(t)^3} + \frac{F(t) A(t)^2}{4} \right)\, .
\end{eqnarray}
By eliminating the unknown function $F(t)$ from the above equations, one 
can impose a mixed (Robin) boundary condition at the matching surface between 
the PN and BHP spacetimes,
\begin{equation}
\label{eq:bc}
\frac{\partial \Psi_{(\rm e)}(t)}{\partial r_*} = \left(\frac{A(t)-2M}{A(t)+2M}
\right)\left(\frac 6{A(t)} \Psi_{(\rm e)}(t) -\frac{80Q(t)}{A(t)^3}\right) \, .
\end{equation}
This specifies a boundary condition at a given moment in time, but it does
not yet describe how to evolve the matching surface (through evolving the 
reduced-mass motion of the system) and the value of the Zerilli function on
this surface.

One can determine the value of the Zerilli function at later times through
the boundary condition above, and the following additional constraint.
By integrating the Zerilli equation with respect to $u$, one finds that
\begin{eqnarray}
\nonumber
\frac{\partial\Psi_{(\rm e)}(t)}{\partial v} &=& -\frac 14\int_0^{u_s(t)} 
V^{(l=2)}_{\rm (e)}(r)\Psi_{(\rm e)}(u',v)du'\\
&& + \frac{\partial\Psi_{(\rm e)}(0,v)}{\partial v} \, ,
\end{eqnarray}
where we have written $r$ implicitly as a function of $u$ and $v$, and
$u_s(t)$ denotes the value of $u$ at the matching surface for a given time $t$.
Having no ingoing waves forces the second term to be zero, so
\begin{equation}
\label{dPsi_dv}
\frac{\partial\Psi_{(\rm e)}(t)}{\partial v} = -\frac 14\int_0^{u_s(t)} 
V^{(l=2)}_{\rm (e)}(r) \Psi_{(\rm e)}(u',v)du'
\, .
\end{equation}
Because both $\partial \Psi_{(\rm e)}(t)/\partial v$ and 
$\partial\Psi_{(\rm e)}(t)/\partial r_*$ are constrained at the point of the
matching surface, this determines the evolution of $\Psi_{(\rm e)}(t)$ on the 
matching surface.
It is easiest to express the Zerilli function on the matching surface as a 
function of time via $\Psi_{\rm (e)}(t,r_*(t))$.
Then, taking the total derivative,
\begin{equation}
\frac{d\Psi_{(\rm e)}(t)}{dt}\equiv \dot\Psi_{(\rm e)}(t) =
\frac{\partial \Psi_{(\rm e)}}{\partial t} + \frac{dr_*}{dt}
\frac{\partial \Psi_{(\rm e)}}{\partial r_*} \, ,
\end{equation}
using the facts that 
\begin{equation}
\frac{\partial \Psi_{(\rm e)}}{\partial t} = 2\frac{\partial \Psi_{(\rm e)}}
{\partial v} - \frac{\partial \Psi_{(\rm e)}}{\partial r_*} \, ,
\end{equation}
and
\begin{equation}
\frac{dr_*}{dt} = \left(1-\frac{2M}r\right)^{-1}\frac{dr}{dt} \, ,
\end{equation}
along with the relationship $a(t) = 2r(t)$, one can write
\begin{eqnarray}
\label{dPsi_dt}
\dot\Psi_{(\rm e)}(t) &=& 2 \frac{\partial \Psi_{(\rm e)}(t)}{\partial v} \\
\nonumber
&-& \left[1-\frac 12\left(\frac{A(t)+2M}{A(t)-2M}\right)\dot A(t)\right] 
\frac{\partial \Psi_{(\rm e)}(t)}{\partial r_*} \, .
\end{eqnarray}
In the above equation, $\partial \Psi_{(\rm e)}(t)/\partial v$ is given by the
integral of the Zerilli function up to that time, Eq.\ (\ref{dPsi_dv}), and
$\partial \Psi_{(\rm e)}(t)/\partial r_*$ is given by the boundary condition,
Eq.\ (\ref{eq:bc}), at that instant.
As a result, the only term in Eq.\ (\ref{dPsi_dt}) that is not yet fixed is
the expression for $\dot A(t)$.

The term $\dot A(t)$ specifies the time evolution of the reduced mass of the 
binary, which, because it is twice the radius of the matching surface between 
the Schwarzschild and PN metrics, could conceivably evolve via either the PN 
equations of motion or those of a particle in the Schwarzschild spacetime.
We will choose the latter, for the same reason as described in Paper I:
the Schwarzschild Hamiltonian has the advantage that a particle falling 
toward the horizon approaches it exponentially in time, in the limit 
that the particle is near the horizon.
Because we are using this motion to approximate the region inside of which PN
theory holds, we want this space to quickly fall toward the horizon as the
theory begins to converge slowly.
Moreover, the motion should move smoothly toward the horizon (so as not to
introduce high-frequency modes that could escape the black-hole effective
potential).
The PN equations of motion do not have these desirable features; we
consequently favor the point-particle evolution equations in the Schwarzschild
spacetime.

We write the evolution equations for the reduced mass of the system
in their Hamiltonian form.
As in Paper I, we will describe the dynamics of the reduced mass in PN 
coordinates, because at late times, this causes the point particles in the
PN metric to approach the horizon in the external Schwarzschild spacetime 
as the reduced mass of the system does the same.
The equations of motion for the reduced mass, $\mu$, are
\begin{eqnarray}
\label{eom}
\nonumber
\dot A(t) = \frac{\partial H}{\partial p_A(t)}\, , &\quad&
\dot \alpha(t) = \frac{\partial H}{\partial p_\alpha(t)}\, \\
\dot p_A(t) = -\frac{\partial H}{\partial A(t)} \, , &\quad&
\dot p_\alpha(t) = \mathcal F_\alpha(t) \, ,
\end{eqnarray}
where the Hamiltonian of a point particle in the Schwarzschild spacetime is 
given by
\begin{eqnarray}
\label{ham}
&& \frac{H(A(t),p_A(t),p_\alpha(t))}{\mu} = \\
\nonumber
&& \sqrt{ \left(1-\frac{2M}{A(t)} \right)\left[1 + \left(1 - \frac{2M}{A(t)}
\right) \frac{p_A(t)^2}{\mu^2} + \frac{p_\alpha(t)^2}{\mu^2A(t)^2}\right]} \, .
\end{eqnarray}
The radiation-reaction force is given by the derivative of the
radiation-reaction potential with respect to $\varphi$, and it should
be evaluated at the location of the matching region, 
\begin{equation}
\frac{\mathcal F_\alpha(t)}{\mu} = \frac 1\mu \frac{\partial U_N^{(l=2),F}}
{\partial \varphi} = -A^2(t)\sqrt{\frac{15}{2\pi}}\Im[F(t)e^{2i\alpha(t)}] \, ,
\end{equation}
where $U_N^{(l=2),F}$ represents the quadrupole part of the radiation-reaction
potential.
By solving Eq. (\ref{eqZQFb}) for $F(t)$ in terms of $\Psi_{(\rm e)}(t)$ 
[and because $Q(t)$ is proportional to a real amplitude times 
$e^{-2i\alpha(t)}$, see Eq.\ (\ref{eq:quad})], one can write
\begin{equation}
\label{force}
\frac{\mathcal F_\alpha(t)}{\mu} = -\sqrt{\frac{15}{2\pi}}\frac{24}{A(t)}
\Im[\Psi_{(\rm e)}(t)e^{2i\alpha(t)}] \, .
\end{equation}

With the above relationship between the radiation-reaction force and the 
Zerilli function, there is now a complete set of evolution equations for the 
reduced-mass motion of the system, the Zerilli function on the matching
surface, and the Zerilli function in the exterior spacetime.
This system of equations is given by
\begin{eqnarray}
\label{eq:full1}
\dot A(t) &=& \frac{\partial H}{\partial p_A(t)}\, , \quad
\dot \alpha(t) = \frac{\partial H}{\partial p_\alpha(t)}\, , \\
\dot p_A(t) &=& -\frac{\partial H}{\partial A(t)} \, , \\ 
\dot p_\alpha(t) &=& -\mu \sqrt{\frac{15}{2\pi}}\frac{24}{A(t)} 
\Im[\Psi_{(\rm e)}(t)
e^{2i\alpha(t)}]\, ,\\
 \dot\Psi_{(\rm e)}(t) &=& -\frac 12\int_0^{u_s(t)} V^{(l=2)}_{\rm (e)}(r) 
\Psi_{(\rm e)}(u',v)du' \nonumber \\
&& - \left[\left(\frac{A(t)-2M}{A(t)+2M}\right)- \frac{\dot A}2 \right]
\nonumber \\
&& \times \left(\frac{6\Psi_{\rm (e)}(t)}{A(t)} - \frac{80Q(t)}{A(t)^3} \right)\, ,\\
\frac{\partial^2 \Psi_{\rm (e)}}{\partial u\partial v} & = & -
\frac{V_{\rm (e)}^{(l=2)}(r)\Psi_{\rm (e)}(u,v)}{4} \, ,
\label{eq:full2}
\end{eqnarray}
where the Hamiltonian is given by Eq.\ (\ref{ham}), the potential by
Eq.\ (\ref{eq:pot}), and the quadrupole by Eq.\ (\ref{eq:quad}).
By including a radiation-reaction force, we arrived at a set of evolution 
equations that simultaneously evolve the reduced-mass motion of the binary and 
the gravitational waves emitted, taking into account the back action of the
emitted radiation on the reduced-mass motion. 

\subsection{Weak-Field Analytical Solution}

First, we will confirm that our procedure recovers the correct 
Burke-Thorne radiation-reaction potential in the weak-field limit.
If we have an equal-mass binary in a circular orbit at a large separation,
$r_*\approx r\approx R \gg M$, then the leading-order behavior of the
Zerilli equation, Eq.\ (\ref{wave_eq}) is just a wave equation in flat space,
\begin{equation}
\frac{\partial^2 \Psi_{(\rm e)}}{\partial t^2} - 
\frac{\partial^2 \Psi_{(\rm e)}}{\partial R^2}
+ \frac 6{R^2}\Psi_{(\rm e)} = 0 \, .
\end{equation}
If one assumes a product solution $\Psi_{(\rm e)} = e^{i\omega t}\psi(R)$,
then for the radial motion, one must solve the ordinary differential 
equation
\begin{equation}
\frac{d^2\psi}{dR^2} + \omega^2 \psi = \frac 6{R^2}\psi \, .
\end{equation}
The solutions for $\psi/R$ are spherical Hankel functions 
$\psi/R = h_2(\omega A/2) = j_2(\omega A/2) + i n_2(\omega A/2)$, assuming
there are no ingoing waves.
Here $\omega$ corresponds to the gravitational-wave frequency.
We must match this wave-zone solution to the PN near-zone expression for the
Zerilli function given by Eq.\ (\ref{eqZQFb}); additionally, we must
also match the derivative of the Hankel function with the radial derivative
of the PN Zerilli function given in Eq.\ (\ref{eqdPsi_drsb}).

We will write these conditions in the frequency domain, where
\begin{eqnarray}
B(\omega)A h_2(\omega A/2) &=& \frac{8Q(\omega)}{A^2} + 
\frac{F(\omega) A^3}{24}\, ,\\
B(\omega)A h_2'(\omega A/2) &=& -\frac{32Q(\omega)}{A^3}
+ \frac{F(\omega) A^2}{4} \, ,
\end{eqnarray}
and we must solve for the unknown amplitude $B(\omega)$ and the 
radiation-reaction potential $F(\omega)$ in terms of the quadrupole
moment $Q(\omega)$ and the spherical Hankel function $h_2(\omega A/2)$.
Since the matching takes place at very large radii, and, by Kepler's law
$A\omega \sim A^{-1/2}$ for circular orbits, one can expand the Hankel
function in $A \omega/2$.
This allows one to solve for $F$ as a series in $1/A$, whose three lowest
terms are given by
\begin{equation}
F = \frac{16}{5A^3} \omega^2 Q + \frac{8}{25A} \omega^4 Q 
+ i \frac{2}{15} \omega^5 Q + O(A^{-6}) \, .
\end{equation}
The third term is the familiar Burke-Thorne radiation-reaction potential
(written in the time domain, this is proportional to five derivatives of
the quadrupole moment).
The first two terms resemble 1PN and 2PN corrections to the Newtonian
potential in the near zone; however, these terms represent the effects of
time retardation that are needed to match the near-zone solution to an 
outgoing wave solution in the wave zone.
As a result, our method recovers, asymptotically, the expected result.
Consequently, the evolution system, Eqs.\ (\ref{eq:full1}) -- (\ref{eq:full2}),
will also give rise to the correct dynamics in the weak-field limit.

\section{Numerical Method and Results}
\label{sec:numerics}

We begin this section by describing the numerical method that we use to solve
the system of evolution equations, Eqs.\ (\ref{eq:full1}) -- (\ref{eq:full2}).
We then show that the evolution equations give rise to reasonable and
convergent results.
With this established, we compare our waveform with one from a 
numerical-relativity simulation, and we close this section by interpreting 
the spacetime of the hybrid method.

\subsection{Numerical Methods and Consistency Checks of the Evolution Equations}
\label{sec:implementation}

\begin{figure}
\includegraphics[width=0.9\columnwidth]{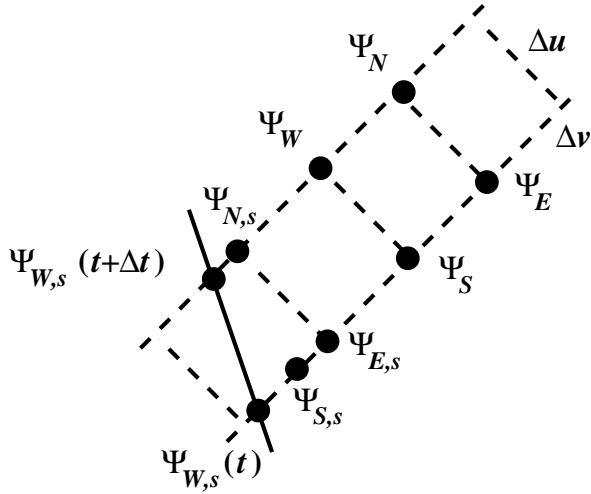}
\caption{A diagram of how we discretize and evolve the Zerilli function.
The dots represent the Zerilli function evaluated at the grid points (the
points of intersection of the dashed lines), and the solid black line is
the matching surface.
For all points except those adjacent to the solid line, one can use Eq.\ 
(\ref{eq:psi_north}) directly to numerically evolve the Zerilli function.
Near the solid line, one can use the same procedure as described in 
Eq.\ (\ref{eq:psi_north}), except that one must interpolate the Zerilli 
function to the point $\Psi_{S,s}$ to use the same procedure.
Further detail is given in the text of this section.}
\label{fig:num_diagram}
\end{figure}

Because the set of evolution equations Eqs.\ (\ref{eq:full1}) -- 
(\ref{eq:full2}) has a somewhat unusual form, we describe our numerical method 
in detail, and we present a few basic checks of the waveform and its 
convergence.
To find the field outside the matching surface, we use the same method as 
that described in Paper I, a second-order accurate, characteristic method.
If we define the following points on the discretized grid 
(see the portion on the right, away from the solid line, in Fig.\
\ref{fig:num_diagram}):
\begin{eqnarray}
\Psi_N = \Psi_{\rm (e)}^{l,m}(u+\Delta u,v+\Delta v)\, , & &
\Psi_W = \Psi_{\rm (e)}^{l,m}(u+\Delta u,v)\, ,\nonumber \\
\Psi_E = \Psi_{\rm (e)}^{l,m}(u,v+\Delta v)\, , & & 
\Psi_S = \Psi_{\rm (e)}^{l,m}(u,v)\, , 
\end{eqnarray}
then discretizing  Eq.\ (\ref{eq:full2}), one can solve for $\Psi_N$
in terms of the other three discretized points and the potential:
\begin{eqnarray}
\nonumber
\Psi_N &=& \Psi_E + \Psi_W - \Psi_S - \frac{\Delta u\Delta v}{8} 
V^l_{\rm (e)}(r_c) (\Psi_E + \Psi_W)\\
&& + O(\Delta u^2 \Delta v, \Delta u \Delta v^2) \, .
\label{eq:psi_north}
\end{eqnarray}
Here $r_c$ is the value of $r$ at the center of the discretized grid,
$(u+\Delta u/2,v+\Delta v/2)$.

We must evolve this partial differential equation simultaneously with the
five ordinary differential equations describing the Zerilli function on 
the matching surface and the surface's position, because all these equations 
are coupled together.
We solve the ordinary differential equations using a second-order accurate
Runge-Kutta method.
As in Paper I, the Zerilli function along the matching surface does not always 
lie on the uniform grid in the $u$-$v$ plane, and we must be careful when 
finding the Zerilli function at grid points adjacent to the matching surface.
For example, at a given value of $u$ along the discretized grid, it is
rare that the Zerilli function on the matching surface, denoted by
\begin{equation}
\Psi_{W,s}(t) = \Psi_{\rm (e)}(u_s(t),v_s(t))\, .
\end{equation}
will actually fall along a grid point
(see the left side of Fig.\ \ref{fig:num_diagram} near the solid line).
Similarly, when evolving the discretized version of Eqs.\ (\ref{eq:full1}) -- 
(\ref{eq:full2}), it is again unlikely that the Zerilli function along the 
matching surface at the next value of $u$ (advanced by one unit of $\Delta u$),
\begin{equation}
\Psi_{W,s}(t+\Delta t)=\Psi_{\rm (e)}(u_s(t)+\Delta u,v_s(t+\Delta t))\, ,
\end{equation}
will fall at a grid point or even at the same value of $v$ as the previous 
earlier value of the Zerilli function, $\Psi_{W,s}(t)$.

To be able to use Eq.\ (\ref{eq:psi_north}) to find the Zerilli function
at $u=u_s(t)+\Delta u$ for the next grid point in $v$ (which we denote by  
$\Psi_{N,s}$), we must interpolate the Zerilli function at fixed $u=u_s(t)$ 
to the same value of $v=v_s(t)$ as $\Psi_{W,s}(t+\Delta t)$.
We will label this point by
\begin{equation}
\Psi_{S,s} =\Psi_{\rm (e)}(u_s(t),v_s(t+\Delta t)) \, .
\end{equation}
As in Paper I, this interpolation does not influence the convergence of the
algorithm when done with cubic interpolating polynomials.
With the value of the Zerilli function at $u=u_s(t)$ and the nearest grid
point in $v$ (which we will call $\Psi_{E,s}$), one can then find the
point $\Psi_{N,s}$ using  Eq.\ (\ref{eq:psi_north}), where 
$\Psi_E$, $\Psi_W$, and $\Psi_S$ are replaced by $\Psi_{E,s}$,
$\Psi_{W,s}(t+\Delta t)$ and $\Psi_{S,s}$, respectively.

As a final note on the numerical methods, we point out that in the evolution 
equation for the Zerilli function on the matching surface, Eq.\ 
(\ref{dPsi_dt}), the term $\partial \Psi_{(\rm e)}(t)/\partial v$ involves an 
integral of the Zerilli function times the potential, Eq.\ (\ref{dPsi_dv}).
Explicitly evaluating this integral adds to the computational expense
significantly, so we compared the value of 
$\partial \Psi_{(\rm e)}(t)/\partial v$ obtained through performing the 
integral with the value found from evaluating 
$\partial \Psi_{(\rm e)}(t)/\partial v$ numerically using a fourth-order
finite-difference approximation of the derivative, calculated from the Zerilli
function in the adjacent exterior BHP spacetime.
Since the two agreed to within the numerical accuracy of our solution, we
used the finite-difference approximation of 
$\partial\Psi_{(\rm e)}(t)/\partial v$ in our numerical evolutions.

\begin{figure}
\includegraphics[width=0.95\columnwidth]{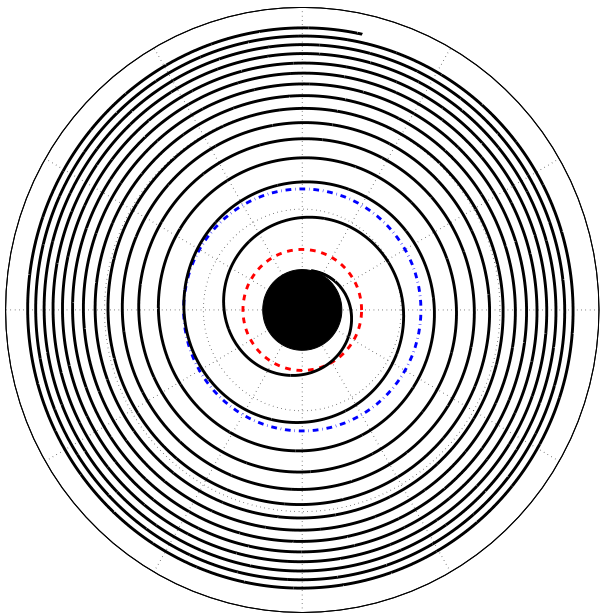}
\caption{ 
In black, the trajectory of the reduced-mass motion of the 
binary, in the PN coordinate system.
The blue (dark) dotted and dashed circle shows the Schwarzschild ISCO, and 
the red (light) dashed circle depicts the light ring of the Schwarzschild 
spacetime.
The large filled black circle represents the horizon.
One can see that the binary plunges soon after it reaches the ISCO of the
exterior Schwarzschild spacetime.}
\label{fig:traj}
\end{figure}

We now examine a few consistency checks of the numerical solutions to the 
system of evolution equations, Eqs.\ (\ref{eq:full1}) -- (\ref{eq:full2}).
In Fig.\ \ref{fig:traj}, we show, in black, the trajectory of the reduced 
mass of the binary in the PN coordinates.
On this same figure, we have depicted the Schwarzschild black hole by a 
filled black circle, the light ring of this black hole by a red (light) dashed 
circle, and the innermost stable circular orbit (ISCO) by a blue (dark) dashed 
and dotted circle.
One can see that the radiation-reaction force causes the matching region
to adiabatically inspiral, until it approaches the ISCO.
Once at the ISCO, it begins plunging more rapidly toward the light ring,
and then falls past the light ring and asymptotes to the horizon of the final 
black hole.

The initial conditions of this evolution correspond to a binary with a PN
separation of $A(0)=14$ in a circular orbit, with no ingoing gravitational 
waves from past-null infinity, and with the radiation-reaction force initially
set to zero.
We do not let the radiation-reaction force enter into the dynamics (thereby 
holding the binary at a fixed separation) until we have a stable estimate
of the force.
At this point, we include the radiation-reaction force (thereby letting the 
binary begin its inspiral).
To minimize eccentricity, we introduce a small change in the radial momentum 
$p_A(0)$ that corresponds to the radial velocity of a PN binary at that 
separation.
Explicitly, we find this value of $p_A(0)$ by solving
\begin{equation}
\dot A(0) = \left.\frac{\partial H}{\partial p_A(t)}\right|_{t=0} 
= -\frac{16}5\frac{M^3}{A(0)^3} \, ,
\end{equation}
(see, e.g.\ \cite{Blanchet}), while assuming that $p_\alpha(0)$ continues to
have the value for circular orbits
\begin{equation}
p_\alpha(0) = \frac{MA(0)}{\sqrt{A(0)/M-3}} \, .
\end{equation}
This is necessary to make the orbit as circular as possible once the binary 
begins to inspiral.
We do not show the initial few orbits before we include the radiation-reaction
force, and we denote the zero of our time to be the moment when we let the
radiation-reaction force begin acting on the binary.

\begin{figure}
\includegraphics[width=0.95\columnwidth]{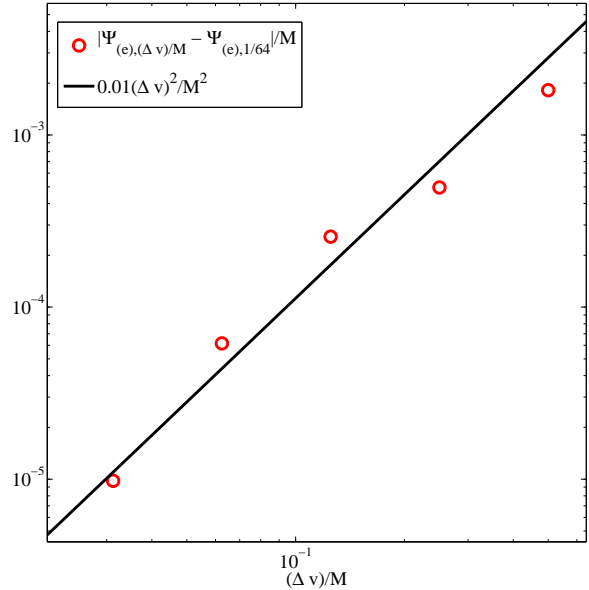}
\caption{
The $L^2$ norm of the Zerilli function at a given
resolution, $\Delta v/M$, minus the Zerilli function at the highest
resolution, $(\Delta v)/M = 1/64$, which we denote by 
$|\Psi_{{\rm (e)},(\Delta v)/M} - \Psi_{{\rm (e)},1/64}|/M$.
We also include a power law proportional to $[(\Delta v)/M]^2$ to indicate
the second-order convergence of our result.}
\label{fig:converge_plot}
\end{figure}

We also calculate the Zerilli function corresponding to these initial 
conditions, as a function of increasing numerical resolution.
In Fig.\ \ref{fig:converge_plot}, we show that the Zerilli function at large 
constant $v$, does converge in a way that is consistent with the 
second-order-accurate code we are using.
We show the $L^2$ norm of the difference between the Zerilli function at
a given resolution, which we denote $\Psi_{{\rm (e)},(\Delta v)/M}$ and the
highest resolution, $(\Delta v)/M = 1/64$, which we denote by
$\Psi_{{\rm (e)},1/64}$.
The $L^2$ norm, therefore, we write as
$|\Psi_{{\rm (e)},(\Delta v)/M} - \Psi_{{\rm (e)},1/64}|$, and we normalize
this by the number of data points in the evolution, and the mass.
We also include a power law, proportional to $[(\Delta v)/M]^2$, which
indicates the roughly second-order convergence of the waveform.

\begin{figure}
\includegraphics[width=0.95\columnwidth]{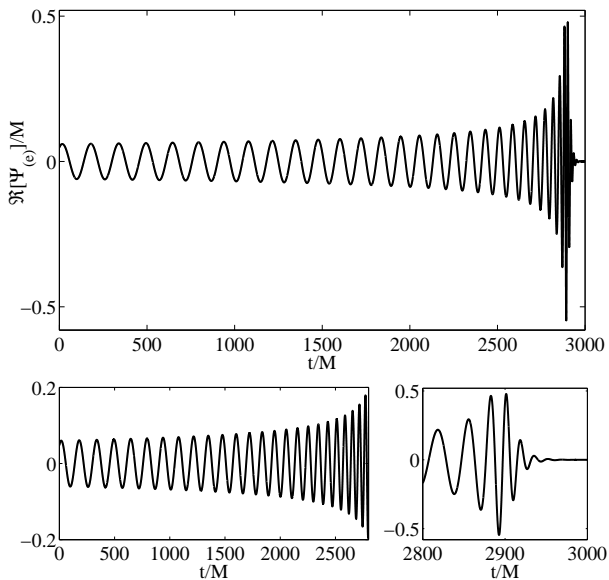}
\caption{The top panel shows the real part of the Zerilli function throughout 
the entire evolution, extracted at large constant $v$.
The bottom-left panel displays the early part of the same Zerilli function, 
and the bottom-right zooms in to the merger and ringdown portions of the 
function.
Because only a factor of $\sqrt 6$ differentiates the Zerilli function from $r$ 
times the waveform, this can be thought of as the waveform as well.}
\label{fig:even_wave}
\end{figure}

We then plot the real part of the Zerilli function extracted at large constant 
$v$, for the highest resolution $(\Delta v)/M = 1/64$, in Fig.\ 
\ref{fig:even_wave}.
The top panel depicts the Zerilli function throughout the full evolution.
Because it is difficult to see the slow increase of the amplitude and frequency
during early times and the smooth transition from inspiral to merger and 
ringdown at late times, we highlight the early stages of the inspiral in the 
lower-left panel, and we depict the merger and ringdown in the lower-right 
panel.
Because $\sqrt 6 \Psi_{\rm (e)} = r(h_+ - i h_\times)$, for the $l=2$ modes
at large $r$ [see Eq.\ (\ref{h_plus})], the Zerilli function is essentially 
identical to the gravitational waveform.
From this one can see the hybrid method produces a smooth 
inspiral-merger-ringdown waveform.
Because the hybrid waveform has the correct qualitative features of a
full inspiral-merger-ringdown waveform, it is natural to ask how well it
could match a numerical-relativity waveform.
We, therefore, turn to this question in the next section.

\subsection{Comparison with Numerical Relativity}

In this section, we will first discuss how well the waveform compares with
a similar waveform from numerical-relativity simulations.
The first part of the section is devoted to showing how we can make small
modifications to the hybrid procedure to make the phase agree well with
that of a numerical-relativity waveform during inspiral (though the comparison
of the amplitudes is less favorable).
The second part of this section describes why the hybrid method, in its current
implementation, does not agree well with numerical-relativity simulations 
during the merger and ringdown phases.
The reason for the discrepancy during the late stages of the waveform is 
well understood (the background spacetime of the hybrid method is 
Schwarzschild, whereas the final spacetime of the numerical simulation is 
Kerr) and could be improved by modifications to the hybrid method.

\subsubsection{Agreement of the Waveforms during Inspiral}

We will briefly describe a small change to the hybrid method that leads to a 
waveform whose phase agrees well with a numerical-relativity waveform during
the inspiral part.
We will continue to find the Zerilli function through the procedure describe
in Sec.\ \ref{sec:rr} using the leading-order expression for the Newtonian 
potential (and thus also the leading-order radiation reaction).
We note, however, that when we took the derivative of the Zerilli function
on the matching surface with respect to $r_*$, Eq.\ (\ref{eqdPsi_drsb}), we 
kept the factor of $(1-2M/r)$.
This is reasonable, physically, because, although the Zerilli function
itself may approach a constant on the horizon, its derivative with respect
to $r_*$ should vanish.
Conversely, if the derivative of the Zerilli function did not vanish, then
that could correspond with a perturbation that diverges on the horizon.
Nevertheless, because the boundary condition only takes into account the 
leading Newtonian expressions, the overall factor of $(1-2M/r)$ is a higher PN 
correction, from the point of view of the interior PN spacetime.
We, therefore, are justified in dropping this term in our leading Newtonian
treatment, and we find the agreement between numerical relativity and 
the hybrid method is helped by this.
It is likely that further adjustments will lead to even better results, though
a systematic study of this is beyond the scope of this initial exposition.

The modification above results in only a small change to 
Eq.\ (\ref{eqdPsi_drsb}), 
\begin{equation}
\frac{\partial \Psi_{(\rm e)}(t)}{\partial r_*} =
-\frac{32Q(t)}{A(t)^3} + \frac{F(t) A(t)^2}{4} \, ,
\end{equation}
and it also alters the boundary condition, Eq.\ (\ref{eq:bc}) of
Sec.\ \ref{sec:rr}, 
\begin{equation}
\frac{\partial \Psi_{(\rm e)}(t)}{\partial r_*} =
\frac {6}{A(t)} \Psi_{(\rm e)}(t) -\frac{80Q(t)}{A(t)^3} \, .
\end{equation}
With the exception of these two equations and the fact that we begin the
evolution from a larger initial radius, $A(0)=15.4$, we evolve the new
system of equations in exactly the same way as that described in detail in 
Sec.\ \ref{sec:implementation}.

\begin{figure}
\includegraphics[width=0.95\columnwidth]{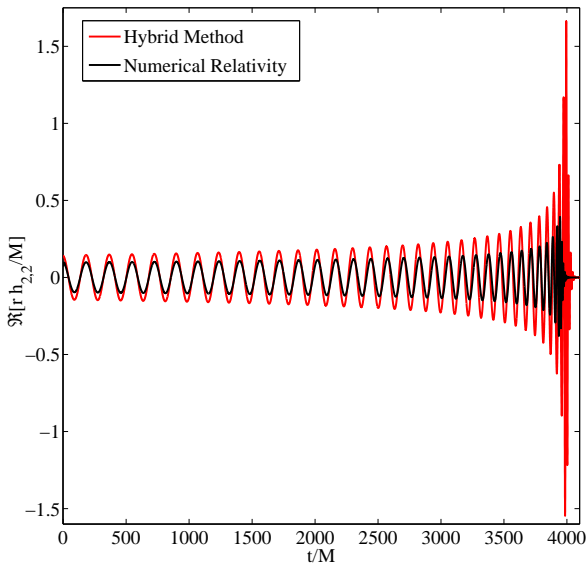}
\caption{In black is the real part of the $l=2$, $m=2$ mode of a 
numerical-relativity waveform, whereas in red (gray) is the equivalent
quantity from the approximate method of this paper.
The agreement of the waveforms' phases is quite good throughout the entire 
inspiral, although the amplitudes differ.
The approximate and numerical-relativity waveforms differ during ringdown,
because the approximate method uses a black-hole with zero spin, whereas the
final black hole in the numerical-relativity simulation has considerable spin.}
\label{fig:hcompare}
\end{figure}

For our comparison with a numerical-relativity waveform, we use the
$l=2$, $m=2$, mode of the waveform from an equal-mass, nonspinning,
black-hole binary described in the paper by Buonanno et al.\ \cite{Buonanno2}.
In this simulation, the black holes undergo 16 orbits before they merge,
and the final black hole rings down.
We plot the numerical-relativity waveform in black in Fig.\ \ref{fig:hcompare},
and we show the equivalent waveform from our approximate method in red (gray).
Recall that the $l=2$ modes of the Zerilli function are related to
the waveform by 
\begin{equation}
\sqrt 6 \Psi_{\rm (e)} = r(h_+-i h_\times) 
\end{equation}
[see Eq.\ (\ref{h_plus})].
Although the amplitudes of the waveforms do not agree exactly, the fact that 
the phases match so well throughout the entire inspiral is noteworthy.
The approximate waveform completes one more orbit than the numerical-relativity
one, and the ringdown portions differ as well.
This is not too surprising, however, since the final black hole in the
numerical-relativity simulation is a Kerr black hole with dimensionless
spin $\chi\approx 0.7$ (see, e.g., Scheel et al.\ \cite{Scheel}), whereas our
ringdown takes place around a Schwarzschild (nonspinning) black hole.

\subsubsection{Differences in the Instantaneous Frequency during Merger
and Ringdown}

\begin{figure}
\includegraphics[width=0.95\columnwidth]{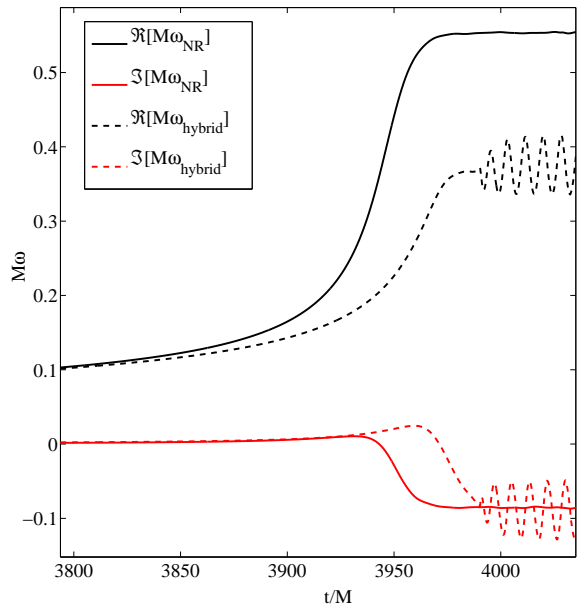}
\caption{ 
The solid curves are the instantaneous frequency 
[see Eq.\ (\ref{eq:frequency})] of the numerical-relativity waveform; the black
curve is the real, oscillatory part and the red (gray) curve is the imaginary,
decaying part.
The black, dashed curve and the red (gray) dashed curve are the real and
imaginary parts, respectively, of the frequency for the hybrid method.
The frequencies agree quite well during the inspiral, but at late times
they begin to differ.
The qualitative transition from inspiral to merger and ringdown is similar,
but the final quasinormal-mode frequencies that the waveforms approach differ,
because the numerical-relativity simulation results in a Kerr black hole
of dimensionless spin $\chi=0.7$, whereas the hybrid waveform is generated
on a Schwarzschild background.
The oscillations in the hybrid waveform arise from the interference of
positive- and negative-frequency modes that can arise in a Schwarzschild
background, as explained in the text of this section.}
\label{fig:frequency}
\end{figure}

The discrepancy between the two waveforms at late times in Fig.\ 
\ref{fig:hcompare} is most evident in the instantaneous frequency, often 
defined as
\begin{equation}
M\omega = i\frac{\dot \Psi_{\rm (e)}}{\Psi_{\rm (e)}} \, ,
\label{eq:frequency}
\end{equation}
where $\Psi_{\rm (e)}$ is the Zerilli function measured at large $r$.
We calculate this frequency for both the hybrid and the numerical-relativity
waveforms, and we show the real and the imaginary parts (the oscillatory
and damping portions, respectively) in Fig.\ \ref{fig:frequency}.
The numerical-relativity waveform was offset from zero at late times by
a small constant of order $10^{-4}$.
We subtracted this constant from the waveform to find the instantaneous
frequency; otherwise, when the amplitude of the waveform becomes comparable
to this constant, there are spurious oscillations in the frequency as it
becomes dominated by this constant offset.
The hybrid waveform needed no modification.

Solid curves depict the instantaneous frequency of the numerical-relativity
waveform in Fig.\ \ref{fig:frequency}; the real (oscillatory) part is the
black curve and the imaginary (decaying) part is the red (gray) curve.
Similarly, the black dashed curve is the real part of the instantaneous frequency
of the hybrid method, and the red (gray) dashed curve is its imaginary part.
The hybrid and the numerical-relativity frequencies are in very good agreement
for the inspiral up until the late stages highlighted here.
The numerical-relativity waveform quickly transitions after the plunge and
merger to the least-damped $l=2$, $m=2$ quasinormal-mode frequency and decay
rate for a Kerr black hole of final dimensionless spin equal to roughly
$\chi\approx 0.7$ (see, e.g., \cite{Leaver}).
The frequency of the hybrid waveform, however, undergoes a similar qualitative
transition, but it approaches the least-damped $l=2$, $m=2$ ringdown frequency
of a non-spinning black hole (the background of the hybrid method).
The hybrid method, however, oscillates around this value with a frequency that
is proportional to twice the frequency of this least-damped, $l=2$, $m=2$
quasinormal mode.

The origin of this oscillation is simple and, in fact, was explained by
Damour and Nagar \cite{Damour}.
For each $l$ and $m$, there are quasinormal modes with both positive and
negative real parts, which both have a negative decay rate.
For a Schwarzschild black hole, the decay rates are the same and the
real frequencies are identical, but have the opposite sign.
For a Kerr black hole, however, the positive-frequency modes have a lower
decay rate than the negative-frequency modes (and the positive frequency is
larger in absolute value than the negative frequency is).
While a counter-clockwise orbit will tend to excite predominantly the mode
with a positive real part, it can also generate the negative real-frequency
mode as well.
In the hybrid waveform, because the background is Schwarzschild, the positive-
and negative-frequency modes decay at the same rate, and they can interfere
to make the oscillations at twice the positive real frequency.
In the numerical-relativity waveform, however, the difference of the
frequencies and decay rates prevents this from happening.

\subsection{Interpreting the Hybrid Waveform and Spacetime}

\begin{figure}
\includegraphics[width=0.95\columnwidth]{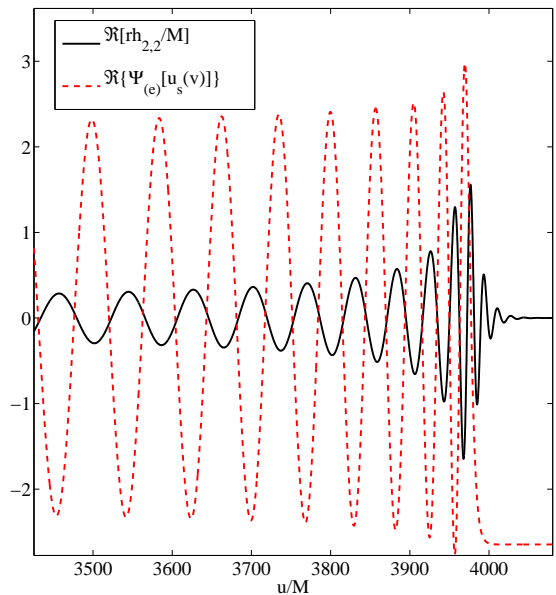}
\caption{The real part of the Zerilli function on the 
matching surface, the red (gray) dashed curve, and the real part of
the gravitational waveform, proportional to the real part of the Zerilli
function at large $v$, (the black solid curve).
The two functions are nearly out-of-phase for the inspiral, and the wave
propagates more or less directly out.
During the merger, they begin to lose this phase relationship, and during 
ringdown the Zerilli function on the matching surface becomes constant. 
This implies that the ringdown waveform is due just to the waves scattered
from the potential, as also illustrated in Fig.\ \ref{fig:field}.}
\label{fig:boundary}
\end{figure}

\begin{figure*}
\includegraphics[width=0.675\textwidth]{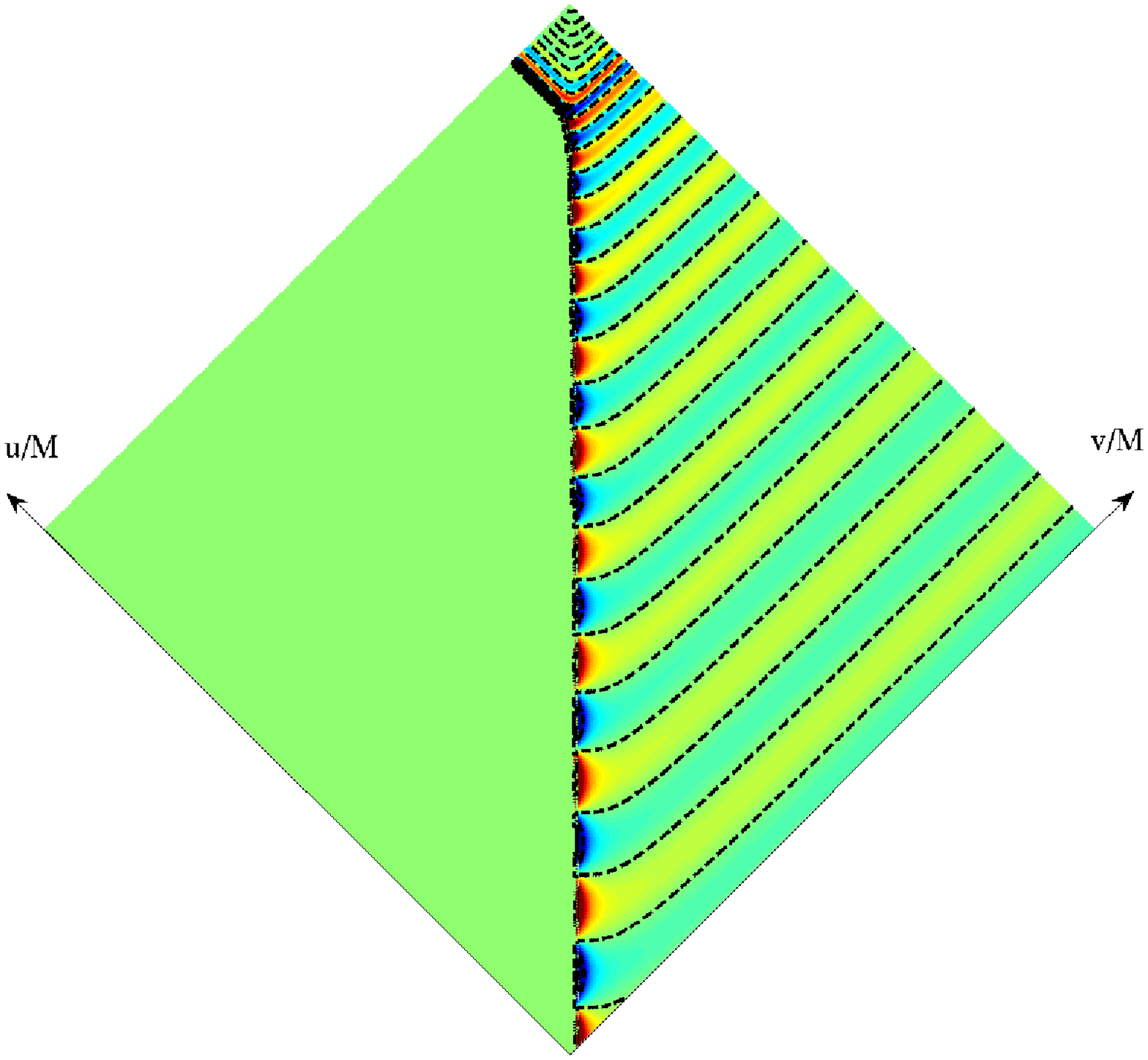} 
\caption{A contour-density plot of the real
part of the Zerilli function for the evolution discussed in this section.
We only show the last few orbits of the inspiral, followed by the merger
and ringdown.
In this spacetime diagram, time runs up, $r_*$ increases to the right, and
the coordinates $u$ and $v$ run at 45 degree angles to the two.
The line that starts at nearly constant $t$ and evolves to a line of
constant $v$ is the matching surface, and to the left of this line, the solid
green (gray) region is the interior PN region (where we do not show any metric 
perturbations).
The exterior is the BHP region, where we show the Zerilli function.
During inspiral, the Zerilli function propagates out almost directly,
and it oscillates between positive, yellow (light gray) colors, and negative,
light blue (darker gray).
Black dashed contour curves are used to highlight this oscillation.
As the reduced mass of the binary plunges into the potential during merger, 
the amplitude and frequency of the radiation increases, but it promptly rings 
down to emit little radiation, in the upper green (gray) diamond of the 
diagram.
There are black dashed contours here as well to indicate that there is still
oscillation, even though it is exponentially decaying (and hard to see through
the color scale).}
\label{fig:field}
\end{figure*}

Since the phase during inspiral agrees so well, and because the transition from
inspiral to merger and ringdown is qualitatively similar, this leads one to
wonder to what extent the hybrid approach may also be a useful tool for  
generating gravitational-wave templates for gravitational-wave searches.
To capture the correct ringdown behavior, the hybrid method would need to be 
extended to a Kerr background; however, it is likely that calibrated approaches using the effective-one-body method (see, e.g., \cite{Buonanno2}) or 
phenomenological frequency-based templates (see, e.g., \cite{Ajith}) will be 
more efficient for these purposes.
The hybrid approach, as described here, will likely be more helpful as a
model of how the near-zone motion of the binary connects to different portions
of the gravitational waveform.

As an example of this, we show the real part of the gravitational waveform at 
large $v$, the black solid curve, and the corresponding value of the Zerilli 
function on the matching surface, the red (gray) dashed curve in Fig.\ 
\ref{fig:boundary}.
Interestingly, the Zerilli function on the matching surface and that extracted 
at large constant $v$ are roughly out-of-phase with one another during the 
inspiral; namely, along a ray of constant $u$, the Zerilli function undergoes 
nearly one half cycle as it propagates out to infinity.
This feature is also visible in Fig.\ \ref{fig:field}, but it is harder
to discern there.
This behavior holds through inspiral up to the beginning of the merger.
During the merger, however, the two transition away from the out-of-phase
relationship, before the Zerilli function on the matching surface becomes a 
constant during the ringdown (when the reduced mass of the binary falls toward
the horizon along a line of constant $v$).

This change in phasing between the Zerilli function on the matching surface and
that at large $v$ (along a line of constant $u$) allows one to give an
interpretation to the different parts of the waveform.
The inspiral occurs when the waveform propagates out directly, but nearly 
out-of-phase with the matching surface.
The merger is the smooth, but brief, transition during which the phase 
relationship between the matching surface and the waveform evolves, and
the ringdown is the last set of waves that are disconnected from the behavior
on the surface (they are the scattered waves from the potential barrier).

We also show in Fig.\ \ref{fig:field} a contour-density plot of the real part 
of the Zerilli function in the $u$-$v$ plane during the last few orbits of 
inspiral, the merger, and the ringdown (for the evolution discussed in this
section).
This is a spacetime diagram, where time runs up, and the radial coordinate,
$r_*$ increases to the right.
The matching surface is the dark timelike curve running up that turns to a 
line of constant $v$ at the end.
The region to the left of the surface, the solid green (gray) is the interior
PN region, but we do not show the metric perturbation in this region.
On its right is the BHP region, where we show the Zerilli function colored so 
that blue colors (dark gray) are negative and red colors (light gray) are 
positive.
Away from the matching surface, the Zerilli function oscillates between yellow 
(light gray) and light blue (darker gray) for several orbits before inspiral.
Each oscillation is bounded between a black, dashed contour curve.
As the reduced mass of the binary plunges toward the horizon, the outgoing 
waves increase in frequency and amplitude, which is how we describe the 
transition from the inspiral to the merger phase.
The merger phase is short, and the black hole rings down (leading to very
little gravitational-wave emission in the top corner of the diagram).
As the reduced mass of the system approaches the horizon, there is a small 
wavepacket of ingoing radiation that accompanies it.

We close this section with one last observation.
If we were to plot the equivalent quantities to those in Figs.\ 
\ref{fig:boundary} and \ref{fig:field} for the evolution in Sec.\ 
\ref{sec:radrxn}, then one would see that the Zerilli function on the matching 
surface increases during ringdown instead of approaching a constant.
This does not have any effect on the waveform, because it is a low frequency
change that occurs within the potential barrier, and is hidden from the
region of space outside the potential.
In some sense, it is a strong confirmation of Price's idea that the details
of the collapse will be hidden within the potential barrier.
At the same time, however, this behavior arises from the fact that the
derivative of the Zerilli function with respect to $r_*$ vanishes on the 
matching surface.
When this condition was neglected in this section, it led to a more regular
behavior there.
This suggests that it may be worth while to do a more careful analysis of how
the Zerilli function and its derivatives near the horizon should scale in the
presence of radiation reaction.

\section{Spinning Black Holes, Spin Precession, and the Superkick Merger}
\label{sec:superkick}

In this section, we will incorporate the effects of black-hole spins into
our method, with the aim of understanding the large kick that arises from the 
merger of equal-mass black holes with spins antialigned and in the orbital 
plane (the superkick configuration).
To do this, we will first discuss adding odd-parity metric perturbations to
the results in the previous section.
We will then indicate why spin precession is important in producing large
kicks and discuss two ways of implementing spin precession: the PN equations 
of precession and geodetic precession in the Schwarzschild spacetime. 
In our method, we will use the geodetic-precession approach, and we will
present numerical results for the kick that uses this equation of spin
precession.

\subsection{Odd-Parity Metric Perturbations of Spinning Black Holes}

To incorporate the effects of spin into our model, we will add the
lowest-order metric perturbation arising from using spinning bodies in the
PN metric, as we did in Paper I.
This comes from the metric coefficients
\begin{equation}
h_{0i} = -\frac{2\epsilon_{ijk} S^j_A n^k_A}{R_A^2} - 
\frac{2\epsilon_{ijk} S_B^j n^k_B}{R_B^2} \, .
\end{equation}
Here we use the notation of Paper I, where we label the two bodies by $A$
and $B$.
The new variables $S_A^j$ represent the spin angular momentum of the body,
$R_A$ is the distance from body $A$ and $n_A^k$ is a unit vector pointing
from body $A$.
The variables for body $B$ are labeled equivalently.
Since we will focus on the extreme kick configuration, we will assume the
black holes lie in the $x$-$y$ plane, at positions ${\bf X}_A(t)$ and
${\bf X}_B(t)$ [identical to Eq.\ (\ref{eq:x_pos}) of Sec.\ \ref{sec:rr}],
and that the spins are given by
\begin{equation}
{\bf S}_A(t) = -{\bf S}_B(t) = S (\cos\beta(t),\sin\beta(t),0) \, ,
\label{eq:spins}
\end{equation}
where $S = \chi (M/2)^2$ is the magnitude of the spin, and $\chi$ is the 
dimensionless spin, ranging from zero to one.

Under these assumptions, one can show that the Cartesian components of
the metric coefficients above are
\begin{eqnarray}
h_{0X} &=& -\frac{3SA(t)}{R^3}\sin 2\theta \sin\beta(t) \cos(\alpha(t)-\varphi) 
\, ,\\
h_{0Y} &=& \frac{3SA(t)}{R^3} \sin 2\theta \cos\beta(t) \cos(\alpha(t)-\varphi)
\, ,\\
h_{0Z} &=& \frac{2SA(t)}{R^3}\sin[\alpha(t)-\beta(t)]\\
\nonumber
&& + \frac{6Sa(t)}{R^3}\sin^2\theta \cos(\alpha(t)-\phi) \sin(\beta(t)-\varphi) \, .
\end{eqnarray}
One can then convert the Cartesian components into spherical-polar coordinates
to find that
\begin{eqnarray}
h_{0R} &=& \frac{2SA(t)}{R^3}\sin[\alpha(t)-\beta(t)]\cos\theta \, ,\\
h_{0\theta} &=& -\frac{2SA(t)}{R^2}\sin[\alpha(t)-\beta(t)]\cos\theta \\
\nonumber
&&-\frac{6SA(t)}{R^2}\sin\theta \cos(\alpha(t)-\varphi) \sin(\beta(t)-\varphi) 
\, ,\\
h_{0\varphi} &=&  \frac{6SA(t)}{R^2}\sin^2\theta \cos\theta
\cos(\alpha(t)-\varphi)\\
\nonumber
&&\times \cos(\beta(t)-\varphi) \, .
\end{eqnarray}
As written above, the metric perturbations do not take the form of an
odd-parity vector harmonic, because there is a dipole-like piece in two of
the components.
This can be eliminated by making a gauge transformation, 
\begin{equation}
\xi_0 = -\frac{SA(t)}{R^2} \cos\theta \sin[\alpha(t)-\beta(t)] \, .
\end{equation}
A small gauge transformation produces a change in the metric via
\begin{equation}
\hat h_{\mu\nu} = h_{\mu\nu}-\xi_{\mu,\nu} - \xi_{\mu,\nu} \, ,
\end{equation}
which in this case sets $\hat h_{0R} = 0$.
The remaining terms in the metric can then be expressed in terms of
the odd-parity, vector spherical harmonics,
\begin{eqnarray}
{\bf X}^{2,\pm 2} &=& (X_\theta^{2,\pm 2},X_\varphi^{2,\pm 2}) \\
\nonumber 
&=& \frac 12 \sqrt{\frac{15}{2\pi}}\sin\theta e^{\pm i 2\varphi}
(\mp i, \sin\theta\cos\theta)\, , \\
{\bf X}^{2,0} &=& (X_\theta^{2,0},X_\varphi^{2,0}) = 
-\frac 32 \sqrt{\frac{5}{\pi}}\sin^2\theta \cos\theta (0,1) \, .
\end{eqnarray}
A short calculation shows that
\begin{eqnarray}
(\hat h_{0\theta},\hat h_{0\varphi}) &=& 2\Re \left[ \frac{SA(t)}{r^2} 
\sqrt{\frac{6\pi}5} e^{-i[\alpha(t)+\beta(t)]} {\bf X}^{2,2}  \right] \\
\nonumber
&& - \frac{8SA(t)}{r^2} \sqrt{\frac \pi 5} \cos[\alpha(t)-\beta(t)]
{\bf X}^{2,0} \, .
\end{eqnarray}

As with the even-parity, mass-quadrupole perturbations discussed in the
previous section, we will only be interested in evolving the $m=2$
perturbation (though in this case it is an odd-parity, current-quadrupole
moment).
The reason for this is subtle, and will be clarified in the next section.
Nevertheless, we will mention here that during the merger and ringdown
(when the kick is generated), the spins precess at the orbital frequency
[namely $\dot\alpha(t) = \dot\beta(t)$].
As a result, the $m=0$ part of the perturbations which depend on 
$\alpha(t)-\beta(t)$ become constant, and the only changes in the perturbations
come from changes in $A(t)$.
We mentioned in Sec.\ \ref{sec:rr} that we would also neglect the $m=0$
part of the even-parity perturbations, because it also evolved from 
time variations in $A(t)$, which occur on the time scale of the 
radiation-reaction force (2.5 PN orders below the leading-order orbital 
motion).
Consequently, because we are interested in the behavior of the binary
during merger and ringdown, we can neglect the $m=0$ parts of the odd-parity
metric perturbations for this same reason.
In addition, because we are treating just the $m=\pm 2$ perturbations (and 
the $m=-2$ term is the complex conjugate of the $m=2$ moment),
we will again drop the label $m$ on the perturbations.

Thus, the relevant piece of the gravitomagnetic potential for our calculation
will be
\begin{equation}
w_{(\rm o)} = -\frac{SA(t)}{4R^2} \sqrt{\frac{6\pi}5} e^{-i[\alpha(t)+\beta(t)]} \, ,
\end{equation}
and one can then use Eq.\ (\ref{eq:gauge_invariant_o}) and the fact that
$R = r - M$ to find that the Regge-Wheeler function is (at leading order in 
$r$),
\begin{equation}
\Psi_{(\rm o)} = \frac{2SA(t)}{R^2} \sqrt{\frac{6\pi}5} 
e^{-i[\alpha(t)+\beta(t)]} \, .
\end{equation}
This means that on the matching surface, 
\begin{equation}
\Psi_{(\rm o)} = \frac{8S}{A(t)} \sqrt{\frac{6\pi}5} e^{-i[\alpha(t)+\beta(t)]} \, .
\label{eq:psi_odd_bndry}
\end{equation}
We can then evolve the Regge-Wheeler equation, Eq.\ (\ref{wave_eq}), (with the
odd-parity $l=2$ potential) using Eq.\ (\ref{eq:psi_odd_bndry}) as the 
boundary condition along the matching surface.
We will not take any radiation-reaction effects from the current-quadrupole 
perturbations into account (since they are 1.5 PN orders below the 
leading-order Newtonian radiation reaction of Sec.\ \ref{sec:rr});
as a result, we will evolve the Regge-Wheeler function using the matching 
surface generated by the even-parity, mass-quadrupole perturbations alone.

\subsection{Spin Precession}

Before we discuss the evolution of the Regge-Wheeler and Zerilli functions, 
we will mention an effect that is important for our recovering the 
correct qualitative behavior of the kick in superkick simulations.
This effect was observed by Schnittman et al.\ in \cite{Schnittman} and 
clarified to us by Thorne \cite{Thorne}.
In Schnittman et al.'s discussion of the superkick configuration, the authors
observe that the spins precess in the orbital plane very rapidly during the
merger, approaching the orbital frequency just before the ringdown.
We will give a heuristic argument of why this effect should occur before
we explore two models that produce spin precession (one based on the PN
equations of motion and the other based on geodetic precession in the 
Schwarzschild spacetime).
We will ultimately favor the latter.

\subsubsection{Motivation for Spin Precession}

One can see the need for spin precession from the following simple argument.
Just as the even-parity perturbations gave rise to a waveform that increased
from twice the orbital frequency to the quasinormal-mode frequency during
the merger phase (see Fig.\ \ref{fig:frequency}), so too must the odd-parity
perturbations of the previous section give rise to a part of the waveform that 
transitions from the orbital frequency to the same quasinormal-mode frequency
as the even-parity perturbations.
The quasinormal-mode frequencies are the same, because both the Regge-Wheeler
and Zerilli functions are generated by $l=2$, $m=\pm 2$ perturbations.
Because the Zerilli function is generated by a boundary condition proportional
to $e^{-i2\alpha(t)}$ and the Regge-Wheeler function produced by a
boundary condition that changes as $e^{-i[\alpha(t)+\beta(t)]}$, for the two
perturbations to evolve in the same way, both $\alpha(t)$, the orbital
evolution, and $\beta(t)$, the spin precession, should evolve in identical
ways at the end of merger.
Stated more physically, at the end of merger, the spins should precess at the
orbital frequency.

This rapid precession of the spins was observed by Br\"ugmann et al.\ 
\cite{Brugmann} in their study of black-hole superkicks.
Using a combination of PN spin precession and numerical-relativity data,
they were able to match the precession of the spin in their numerical
simulations.
We will explain in the next section why this worked so well for their 
simulation, but why it will not work as well in the hybrid method.

\subsubsection{Post-Newtonian Spin Precession}

Br\"ugmann et al.\ begin from the well-known spin precession for a binary
(see, e.g., \cite{Kidder}), 
\begin{equation}
\dot {\bf S}_A(t) =  \frac 1{A(t)^3} \left(2+\frac{3M_B}{2M_A} \right)
[{\bf L}_N(t) \times {\bf S}_A(t)]
\label{eq:spin_prec}
\end{equation}
where we just write the leading-order effect from the Newtonian angular 
momentum,
\begin{equation}
{\bf L}_N(t) = \mu \{[{\bf X}_A(t)-{\bf X}_B(t)] \times [\dot{\bf X}_A(t)-
\dot{\bf X}_B(t)]\}
\end{equation} 
The vector $\hat{\bf n}$ is a unit vector from the center of mass.
There is an equivalent equation for the precession of ${\bf S}_B(t)$,
identical to the equation above, under the interchange of $A$ and $B$.
Given the form of the equation above, the magnitude of the spin does not 
change, and the spin precesses about the Newtonian angular momentum 
${\bf L}_N(t)$.
Moreover, Br\"ugmann et al.\ found that for the superkick configuration,
where the spins lie in the plane, precession of the spins does not produce
a large component out of the plane (the $z$ component in this case).

For simplicity, therefore, we will just consider the components of the spin 
in the orbital plane, which, at leading-order, will precess as a result of
coupling to the Newtonian orbital angular momentum.
The Newtonian angular momentum is
\begin{equation}
{\bf L}_N(t) = \mu A(t)^2\dot\alpha(t) \hat{\bf z} \, ,
\end{equation}
where $\dot\alpha(t)$ is the orbital frequency.
With the assumption that $S_A^z = S_B^z = 0$, the spins precess via the equation
\begin{equation}
\dot {\bf S}_A(t) = \frac{7M\dot\alpha(t)}{8A(t)} [\hat{\bf z} \times 
{\bf S}_A(t)] \, ,
\end{equation}
where we have also used the fact that this is an equal-mass binary,
($M_A = M_B = M/2$ and $\mu = M/4$).
Taking the time derivative of Eq.\ (\ref{eq:spins}), we obtain the expression
for the left-hand side of the equation above,
\begin{equation}
\dot {\bf S}_A(t) = \dot\beta(t) [\hat{\bf z} \times {\bf S}_A(t)] \, ,
\end{equation}
Relating the two expressions, we arrive at the equation of spin precession, 
\begin{equation}
\dot\beta(t) = \frac{7M}{8A(t)} \dot\alpha(t) \, .
\label{eq:spin_prec_simp}
\end{equation}
For the hybrid  method, this expression will not lead to the spin-precession 
frequency approaching the orbital frequency, since $A(t) \geq 2M$ for the
entire evolution (and hence, the spin-precession frequency will not even
be half the orbital frequency at its maximum).
In the next section, we will put forward an equation of spin precession based
on geodetic precession in the external Schwarzschild spacetime, which will
have the desired spin-precession behavior.

Before turning to the next section, we address the question of why PN spin
precession worked so successfully for Br\"ugmann et al.
Their initial data begins in a gauge that is identical to the 2PN ADMTT gauge,
and they assume that it continues to stay in that gauge throughout their
evolution.
As a result, they use the puncture trajectories as the positions of the
black holes, and the 2PN ADMTT gauge expressions to relate the momenta of
the black holes to their velocities.
Although the PN equations of spin precession are written in harmonic gauge,
they use the puncture results to calculate these expressions.
This is reasonable, because the harmonic and ADMTT gauge positions do not 
differ much until separations of roughly $A(t) \approx 2M$.
Their puncture separations do reach small values of $A(t) < M$ prior to
merger, and they continue to use the harmonic-gauge spin-precession
formula in this regime (even as the PN approximation starts becoming
less accurate).
This works remarkably well, nevertheless, and, as one can see from Eq.\
(\ref{eq:spin_prec_simp}), when $A(t) \approx 7M/8$, the spins will precess
at the orbital frequency.
Thus, the work of Br\"ugmann et al.\ helps to confirm that the locking of
the orbital and spin-precession frequencies is important in the superkick
merger, but to replicate this effect in the hybrid method will require a
different approach, described below.

\subsubsection{Geodetic Precession in a Schwarzschild Spacetime}

Our approach to spin precession relies on geodetic precession in the
Schwarzschild spacetime, which we review below.
The problem of geodetic precession of a spin on a circular orbit in the 
Schwarzschild spacetime is well understood; its derivation appears in
the introductory text by Hartle \cite{Hartle}, for example.
We will reproduce some of the important elements of the derivation here,
using our notation, however.
One typically starts with the spin 4-vector $S^\mu$ (whose spatial 
components lie in the orbital plane) that travels along a circular
geodesic parametrized by a 4-velocity $u^\mu$.
As usual $u^\mu u_\mu = -1$, and one also imposes the spin-supplementary
condition, $S^\mu u_\mu = 0$.
The components of these two vectors are $\vec S = (S^t, S^r, 0, S^\varphi)$,
and $\vec u = u^t(1,0,0,\dot\alpha(t))$.
Because of the spin-supplementary condition and the normalization of the
four velocity, the components $S^t$ and $u^t$ are not independent
variables.
Thus, when one writes the equation of geodetic precession of the spin 
[Eq.\ (14.6) of Hartle],
\begin{equation}
\frac{dS^\mu}{d\tau} + \Gamma^{\mu}\,_{\rho\nu}S^\rho u^\nu = 0 \, ,
\end{equation}
for circular equatorial orbits, it reduces to two coupled equations for the 
independent variables $S^r(t)$ and $S^\varphi(t)$ [Eqs.\ (14.3a) and (14.3b) of 
Hartle],
\begin{eqnarray}
\dot S^r(t) - [r_s(t)-3M] \dot\alpha(t) S^\varphi(t) & = & 0 \, , \\
\dot S^\varphi(t) + \frac{\dot \alpha(t)}{r_s(t)} S^r(t) & = & 0 \, .
\end{eqnarray}
The dot still refers to derivatives with respect to coordinate time $t$
(not proper time $\tau$).
If we assume that $\dot \alpha(t)$ does not change much over an orbit (which
is true during most of the evolution of the binary, as it changes only due to 
the radiation-reaction force), and we continue to denote the angle of the
spin in the orbital plane by $\beta(t)$, then one can write the solution
to these equations [Eqs.\ (14.16a) and (14.16b) of Hartle] as,
\begin{eqnarray}
S^r(t) & = & S \sqrt{1-\frac{2M}{r_s(t)}} \cos[\alpha(t)-\beta(t)] \, , \\
S^\varphi(t) & = & \frac S{r_s(t)} \sqrt{1-\frac{2M}{r_s(t)}} 
\frac{\dot\alpha(t)}{\dot\alpha(t)-\dot\beta(t)} \nonumber \\
&& \times \sin[\alpha(t)-\beta(t)] \, ,
\end{eqnarray}
where the spin is normalized $S^\mu S_\mu = S^2$, and where 
[Eq.\ (14.15) of Hartle]
\begin{equation}
\dot\alpha(t) - \dot\beta(t) = \sqrt{1 - \frac{3M}{r_s(t)}} \dot \alpha(t) \, .
\end{equation}

Because we only describe the spins with leading-order physics, we will
only keep the leading-order behavior of the spins.
Thus, we will describe the spatial components of the spins by
\begin{equation}
S^r(t) = S \cos[\alpha(t)-\beta(t)] \, , \quad S^\varphi(t) = \frac S{r_s(t)}
\sin[\alpha(t)-\beta(t)] \, ,
\end{equation} 
and we will expand the equation for the evolution of $\beta(t)$ in a Taylor
series up to linear order in $M/r_s(t)$,
\begin{equation}
\dot\alpha(t) - \dot\beta(t) = \left(1 - \frac{3M}{2r_s(t)}\right) 
\dot\alpha(t) \, .
\end{equation}
We ultimately arrive at the expression that we will use to describe spin
precession in our formalism,
\begin{equation}
\dot\beta(t) = \frac{3M}{A(t)} \dot\alpha(t) \, ,
\label{eq:spin_prec_geo}
\end{equation}
since at leading order $A(t) = a(t)$.

Although Eq.\ (\ref{eq:spin_prec_geo}) looks quite similar to the leading-order 
PN spin precession, Eq.\ (\ref{eq:spin_prec_simp}), the former equation 
produces a much stronger spin precession than the latter does.
Not even the next-order PN spin-precession terms will produce such strong
precession (see, e.g., \cite{Faye}).
The equation of spin precession based on geodetic motion takes on more of
the strong-gravity character of the Schwarzschild spacetime.
It states that when a spinning particle orbits near the light ring, its
spin will lock to the its orbital motion.
An effect quite similar to this happens during the merger phase in the 
superkick simulation, as was shown in the work of Br\"ugmann et al., and
which we discussed in the previous section.
In the next section, we will show how this contributes to the large kick
of the superkick simulations.

\subsection{Numerical Results and Kick}
\label{sec:results}

In the first part of this section, we describe how we numerically solve the
Regge-Wheeler equation (we continue to solve the Zerilli equation in the same 
way as described in Sec.\ \ref{sec:numerics}), and we show a representative 
waveform obtained from the Regge-Wheeler function.
We next describe how we calculate the linear-momentum flux and the kick
from the waveforms.
Finally, we close this section by studying the dependence of the kick on the 
initial angle between the spins and the linear momentum of the PN point
particles.
We recover results that are qualitatively similar to those seen in 
full numerical-relativity simulations.

\subsubsection{Numerical Methods and Waveforms}

To calculate the Regge-Wheeler function, and thus the radiated
energy-momentum in the gravitational waves, we first make the following
observation.
Because the odd-parity perturbation of the spins of the black holes
is a 1.5 PN effect, the corresponding radiation-reaction force will also enter
at 1.5 PN beyond the leading-order radiation-reaction force discussed in 
Sec.\ \ref{sec:rr}.
Consequently, we do not take it into account in the leading-order treatment
of the radiation-reaction force.
Moreover, we note that the spin-precession angle, $\beta(t)$, does not
enter into the evolution equations for the reduced mass or for the Zerilli 
function.
As a result, the evolution of $\beta(t)$ and $\Psi_{\rm (o)}$ can be
performed after the evolution of the binary without spin.
In fact, the evolution of $\Psi_{\rm (o)}$ is carried out in the same manner
as that described in Paper I, because the matching surface is driven
by radiation-reaction from the even-parity Zerilli function alone.
Were we to include the radiation reaction arising from the spins, however,
we would need to evolve the equations for $\beta(t)$ and $\Psi_{\rm (o)}$
simultaneously, and in a manner identical to that described in Sec.\ 
\ref{sec:rr}.

\begin{figure}
\includegraphics[width=0.95\columnwidth]{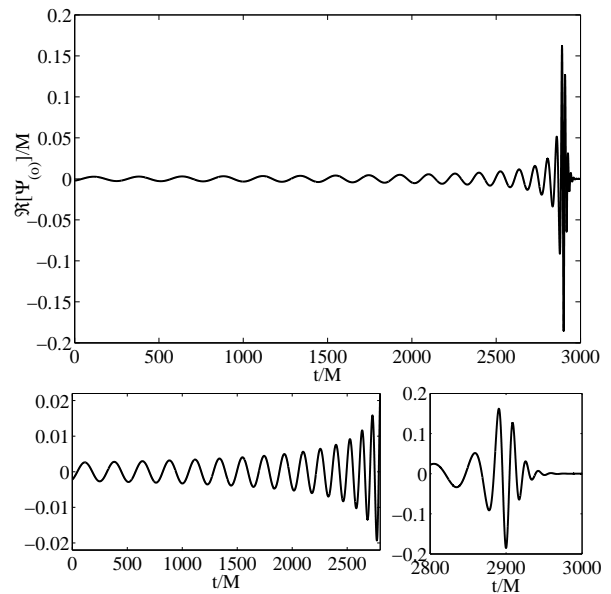}
\caption{The top panel shows the real part of the Regge-Wheeler function 
throughout the entire evolution.
The bottom-left panel just focuses on the early times of inspiral, where the 
Regge-Wheeler function slowly increases in frequency and amplitude because
of the binary's inspiral and the slow spin precession.
In the bottom-right panel, one sees that as the spins begin to precess
near the orbital frequency, the Regge-Wheeler function dramatically increases 
in amplitude and frequency.}
\label{fig:odd_wave}
\end{figure}

Our initial conditions are identical to those described in Sec.\ 
\ref{sec:implementation}, but we will set the dimensionless spin $\chi = 1$,
and let $\beta(0)$ vary over several values from $0$ to $2\pi$, to 
study the influence of the initial angle on the kick.
We first show the real part of the Regge-Wheeler function extracted 
at large constant $v$, in Fig.\ \ref{fig:odd_wave}.
The top panel is the full Regge-Wheeler function, whereas the bottom-left
panel features the early part from the inspiral (so that one can see the 
gradual increase in the amplitude and frequency that comes from the combined 
effects of the binary inspiral, and the increased rate of spin precession).
In the bottom-right panel, we show the merger and ringdown phase, which
is obscured in the top panel.
As the spins start precessing near the orbital frequency during merger,
one can see the rapid growth of the Regge-Wheeler function.

\begin{figure}
\includegraphics[width=0.95\columnwidth]{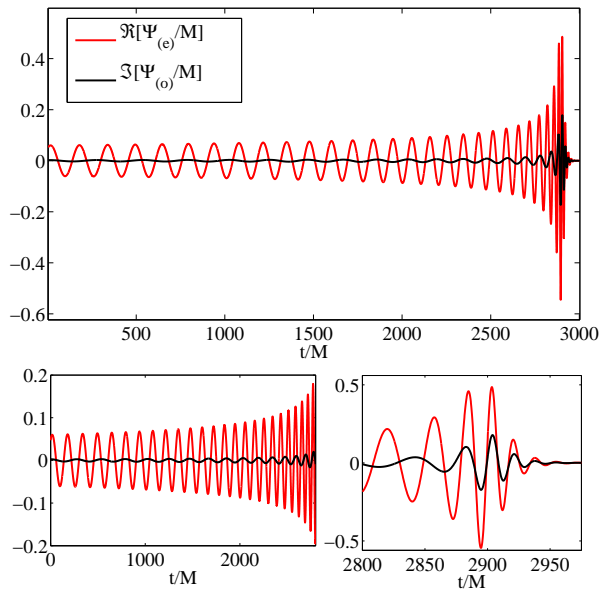}
\caption{
The top panel shows the real part of the Zerilli function, the red (gray)
curve, and the imaginary part of the Regge-Wheeler function, the black
curve, throughout the inspiral, merger, and ringdown.
The bottom-left panel shows only the inspiral, where the Regge-Wheeler 
function is much smaller than the Zerilli function, and oscillates at 
approximately half the frequency.
In the bottom-right panel, during the merger and ringdown, as the spins 
precess near the orbital frequency, the Regge-Wheeler function increases 
in amplitude and frequency, and becomes in phase with the Zerilli function.
This leads to a large kick.}
\label{fig:both_waves}
\end{figure}

To see how this spin precession leads to a large kick, we plot both the even-
and the odd-parity metric perturbations extracted at large constant $v$ in 
Fig.\ \ref{fig:both_waves}.
We show the real part of the Zerilli function, $\Psi_{\rm (e)}$, in red (gray) 
and the imaginary part of the Regge-Wheeler function, $\Psi_{\rm (o)}$, in 
black, for the angle $\beta(0)$ that gives the maximum kick.
As we show below, in Eq.\ (\ref{eq:pzdot}), it is the relative phase of the 
product of these components that is important in producing the kick.
During the early part of the evolution, the Regge-Wheeler function is quite
small and oscillates with roughly half the period of the Zerilli function.
This is difficult to see in the upper panel of the full waveforms in 
Fig.\ \ref{fig:both_waves}, but is more evident in the lower-left panel,
showing just the early parts of the evolution.
In the last orbit before the merger and ringdown (shown in the lower-right 
panel), the spins start precessing rapidly, and, in the case that produces the 
maximum kick, the real part of the even-parity perturbation function, and the 
imaginary part of the odd-parity function oscillate in phase during the merger 
and ringdown.
(For the case with zero kick, the two functions are now out-of-phase
by 90 degrees.)

\subsubsection{Calculation of the Kick}

We now discuss, more concretely, how we calculate the kick emitted in 
gravitational waves.
At radii much larger than the reduced gravitational wavelength,
$r\gg \lambda_{\rm GW}/(2\pi)$, one can relate the gravitational-wave
polarizations $h_+$ and $h_\times$ to the Regge-Wheeler and Zerilli functions
via the expression, 
\begin{eqnarray}
&&h_+ - i h_\times \nonumber \\
& =&\frac{1}{2r}\sum_{l,m} \sqrt{\frac{(l+2)!}{(l-2)!}}\left[
\Psi^{l,m}_{(\rm e)}+i\,\Psi^{lm}_{(\rm o)}\right]{}_{-2}Y_{lm} \, ,
\label{h_plus}
\end{eqnarray}
where ${}_{-2}Y_{lm}$ is a spin-weighted spherical harmonic.
The energy radiated in gravitational waves is typically expressed as
\begin{equation}
\dot P_i(t) = \lim_{r\rightarrow\infty} \frac{r^2}{16\pi} 
\oint n_i |\dot h_+ - i\dot h_{\times}|^2 d\Omega \, ,
\end{equation}
where $n_i$ is a radial unit vector and $d\Omega$ is the area element on a 
2-sphere.
A somewhat lengthy calculation can then show that the momentum flux in the
$z$ direction is given by
\begin{eqnarray}
\dot P_z(t) &=& \frac 1{16\pi} \sum_{l,m} \frac{(l+2)!}{2(l-2)!}
\left[-ic_{l,m} \dot \Psi_{\rm (e)}^{l,m}  \dot{\bar \Psi}_{\rm (o)}^{l,m}
\right. \\
\nonumber
&& \left.+d_{l+1,m}\left(\dot \Psi_{\rm (e)}^{l,m} 
\dot{\bar \Psi}_{\rm (e)}^{l+1,m} \dot \Psi_{\rm (o)}^{l,m} 
\dot{\bar \Psi}_{\rm (o)}^{l+1,m}\right) \right] \, ,
\end{eqnarray}
where $c_{l,m} = 2m/[l(l+1)]$, and $d_{l,m}$ is a constant that also depends
upon $l$ and $m$.
The equations above appear in several sources; these agree with those
of Ruiz et al.\ \cite{Ruiz} [see their Eqs.\ (84), (11), (94), and (43), 
respectively].

In our case, however, we just treat the $l=2$ and $m=\pm 2$ modes of the 
Regge-Wheeler and Zerilli functions, and the momentum flux coming from these 
modes greatly simplifies.
Because the $m=\pm 2$ modes are complex conjugates of one another, we find
that the momentum flux is
\begin{equation}
\dot P_z(t) = \frac 1{\pi}\Im[\dot \Psi_{\rm (e)}\dot{\bar \Psi}_{\rm (o)}]
\, .
\label{eq:pzdot}
\end{equation}
When we discuss the kick velocity as a function of time, we mean that we take
minus the time integral of the momentum flux, normalized by the total mass,
i.e.\
\begin{equation}
v_z^{\rm kick}(t) = -\frac 1M \int_{t_0}^t \dot P_z(t') dt' \, .
\label{eq:v_kick}
\end{equation}
We continue to normalize by the total mass $M$, because numerical-relativity
simulations have shown that it changes only by roughly $4\%$ during a 
black-hole-binary merger (see, e.g., Campanelli et al.\ \cite{Campanelli});
as a result, normalizing by the total mass $M$ will not be a large source of
error.

\subsubsection{Numerical Results of the Momentum Flux and Kick}

\begin{figure}
\includegraphics[width=0.95\columnwidth]{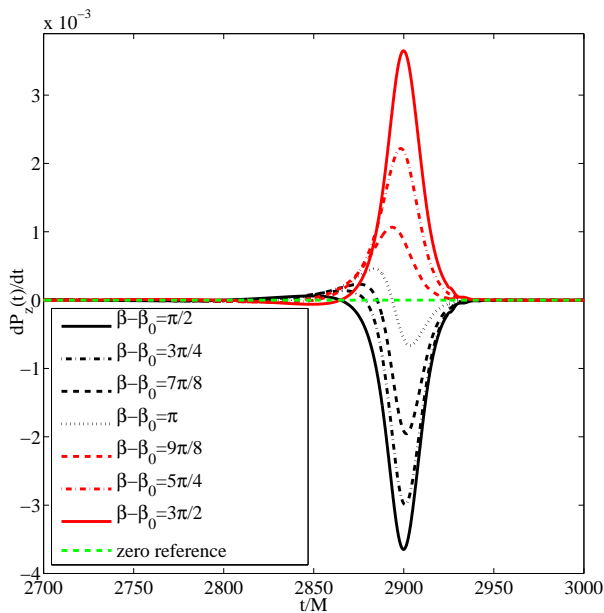}
\caption{
The momentum flux, $\dot P_z(t)$, as function of time, for several values of 
$\beta-\beta_0$, where $\beta_0 = 215\pi/192$ is the value that gives zero 
kick.
We also include a straight green (light gray) dashed line at zero flux to 
indicate how the momentum flux varies around this point.}
\label{fig:dPzdt}
\end{figure}

We now show the results of our numerical solutions for the superkick 
configuration.
We first show in Fig.\ \ref{fig:dPzdt} the momentum flux for several different 
initial angles of the spins, $\beta$.
In the plots, we subtract the value that gives zero kick, which we
denote by $\beta_0 = 215\pi/192$.
While the shape of the pulse of momentum flux has a similar shape to 
that seen in numerical-relativity simulations by Br\"ugmann et al.\ 
\cite{Brugmann}, the absolute magnitude is somewhat larger.

\begin{figure}
\includegraphics[width=0.95\columnwidth]{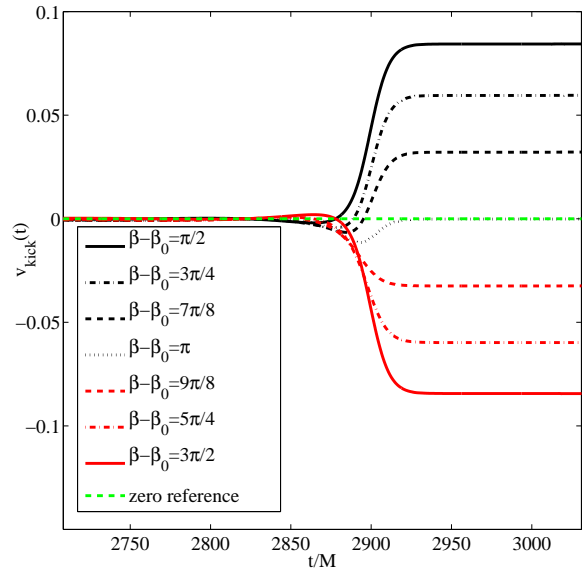}
\caption{
The kick as a function of time, for several different initial angles of the 
spin $\beta-\beta_0$.
There is also a straight green (light gray) dashed line at zero velocity to 
indicate how the kick varies around this point.}
\label{fig:v_kick}
\end{figure}

The increased overall magnitude of the kick becomes more apparent when we plot
$v_{\rm kick}(t)$ in Fig.\ \ref{fig:v_kick}, where $v_{\rm kick}(t)$ is
defined by Eq.\ (\ref{eq:v_kick}).
As one can see, the largest value of the kick is near $0.08$ in dimensionless
units, which is roughly a factor of $6$ times larger than the estimated 
maximum from numerical-relativity simulations at lower dimensionless
spin parameters.
This is largely because the even-parity Zerilli function (proportional to
the waveform) is also significantly larger in amplitude than that of 
numerical-relativity simulations.

\begin{figure}
\includegraphics[width=0.95\columnwidth]{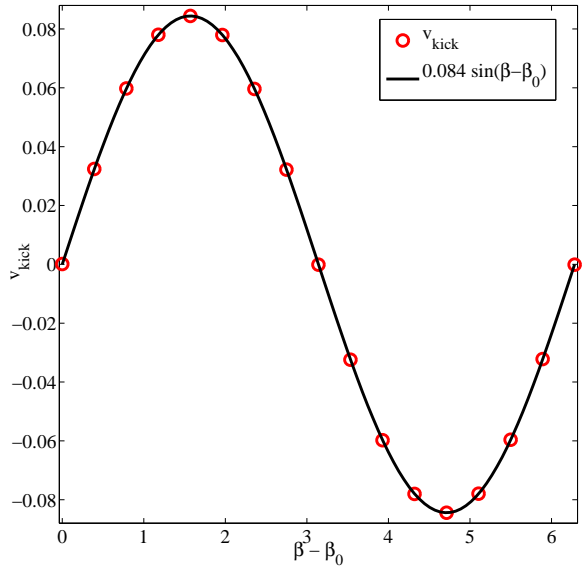}
\caption{
The kick calculated for several initial values of $\beta$,
minus the angle that produces nearly zero kick, $\beta_0 = 215\pi/192$.
The data points are calculated from the numerical evolutions of this section, 
while the solid curve is a sinusoidal fit to the data.
The hybrid model produces a sinusoidal dependence of the kick on the 
initial angle $\beta-\beta_0$ very precisely.}
\label{fig:kick}
\end{figure}

Nevertheless, we then show, in this model, that the kick depends sinusoidally 
upon the initial orientation of the spins, as seen in numerical simulations by
Campanelli et al.\ \cite{Campanelli}.
We plot the final value of the kick, $v_{\rm kick} \equiv v^{\rm kick}_z(t_f)$,
where $t_f$ is the last time in the simulation, as a function of 
$\beta - \beta_0$ in Fig.\ \ref{fig:kick}.
The sinusoidal dependence in our model is exact up to numerical error.
One can see this must be the case from examining the form of our expression
for the momentum flux, Eq.\ (\ref{eq:pzdot}).
Because the evolution equations are not influenced by the orientation of
the spins, then the Zerilli function will be identical for different initial
spin directions.
The Regge-Wheeler function, however, will evolve in the same way, but
because the value on the matching surface is proportional to $e^{-i\beta(t)}$
[see Eq.\ (\ref{eq:psi_odd_bndry})], the different evolutions will also differ 
by an overall phase, $e^{i\beta}$, where $\beta$ is the initial value of the
spin.
Thus, when one takes the imaginary part of product of the Regge-Wheeler and 
Zerilli functions to get the momentum flux in Eq.\ (\ref{eq:pzdot}), one
will have sinusoidal dependence.
(In fact, we could have simply done one evolution and changed the phasing
as described above to find the above results; as a test of our method,
however, we in fact performed the multiple evolutions to confirm this idea.)

We close this section by making the following observation, which may be 
known, though we have not seen in numerical-relativity results.
Since the dependence on $\beta$ of the kick is sinusoidal, then for each 
$\beta$, $\beta - \pi$ gives the same kick and momentum flux pattern,
$\dot P_z(t)$, just opposite in sign.
At the same time, though, because of the sinusoidal dependence there are two 
values that give rise to the same kick in the same direction; however the 
shape of the momentum flux $\dot P_z(t)$ is not the same for these two.
One can see this in Fig.\ \ref{fig:dPzdt}, where the black dotted and dashed
curve and minus the red (gray) dotted and dashed curve give rise to the 
same kick; nevertheless, the pattern of the momentum flux is very different.
A careful study of this would reveal more about how the spins precess
and would be of some interest.

\section{Conclusions}
\label{sec:conclusion}

In this paper, we extended a hybrid method for head-on mergers to treat
inspiralling black-hole binaries.
We introduced a way to include a radiation-reaction force into the hybrid 
method, and this led to a self-consistent set of equations that evolve the 
reduced-mass motion of the binary and its gravitational waves.
Using just PN and linear BHP theories, we were able to 
produce a full inspiral-merger-ringdown waveform that agrees well in phase
(though less well in amplitude) with those seen in full numerical-relativity 
simulations.
Even though the dynamics during inspiral follow the modified dynamics of a
point particle in Schwarzschild rather than the exact dynamics of a black-hole 
binary, the phasing in the waveform agrees well.
Because we assume the background is a Schwarzschild black hole
(rather than a Kerr, the true remnant of black-hole binary inspirals), the
merger and ringdown parts of the hybrid and numerical-relativity waveforms
do not match as well.
Nevertheless, the hybrid method does produce a waveform that is quite
similar to that of numerical relativity.

We also studied spinning black holes, particularly the superkick configuration
(antialigned spins in the orbital plane).
We discussed a method to incorporate spin precession, based on the
geodetic precession of a spinning point particle in the Schwarzschild 
spacetime.
This caused the spins to lock to the orbital motion during the merger and
ringdown, which, in turn, helped to replicate the pattern of the 
momentum flux and the sinusoidal dependence of the merged black hole's 
kick velocity seen in numerical simulations.
Again, because the amplitude of the emitted gravitational waves does not match
that of numerical-relativity simulations, the magnitude of the kick does not
completely agree.
Nevertheless, because the approximate method was able to capture the pattern
of the momentum flux, it gives credence to the idea the locking of the 
spin-precession frequency to the orbital frequency contributes to large 
black-hole kicks.

It would be of interest to extend this approach to see if it can recover
the results of numerical relativity more precisely.
To do this would involve a two-pronged approach:
on the one hand, we would need to include higher PN terms in the metric in the 
interior while using a more accurate Hamiltonian to describe the 
conservative dynamics of the binary (such as the EOB Hamiltonian);
on the other hand, we would need to evolve the perturbations in a Kerr
background.
It would be simpler to choose the Kerr background to have the spin of the
final, merged black hole, but one could also envision evolving perturbations 
in an adiabatically changing Kerr-like background with a slowly varying mass 
and angular momentum parameter that change in response to the emitted 
gravitational waves.
It would be of interest to see if such an approach leads to an estimate of
the spin of the final black hole similar to that proposed by Buonanno, 
Kidder, and Lehner \cite{BKL}.
Incorporating the PN corrections and a new Hamiltonian would be the most 
straightforward improvement, while those involving the Kerr background are 
technically more challenging, and computationally more expensive.

\begin{acknowledgments}

We thank Jeandrew Brink, Tanja Hinderer, Lee Lindblom, Yasushi Mino, 
Mark Scheel, Bel\'a Szil\'agyi, Kip Thorne, Huan Yang and Aaron Zimmerman for 
discussing various aspects of this work with us.
In particular, we acknowledge Scheel for letting us use the waveform from 
the numerical-relativity simulation, Szilagyi for his input on numerical 
algorithms, Mino and Yang for their remarks on the validity of PN and BHP 
theories in the early stage of this work, and Thorne for his constant 
encouragement and for reminding us that spin precession must lock the spins 
to the orbital frequency.
This work is supported by NSF grants No. PHY-0653653 and No.\ PHY-1068881 and 
CAREER Grant No.\ PHY-0956189, and by the David and Barbara Groce start-up fund.
D.N. was supported by the David and Barbara Groce graduate research 
assistantship at Caltech. 

\end{acknowledgments}

\end{document}